\documentclass[]{aa}

\usepackage{graphicx}
\usepackage{natbib}
\usepackage{subfigure}
\usepackage{morefloats}
\usepackage{lscape}

\bibpunct{(}{)}{;}{a}{}{,} 

\begin{document}

\title{Keck spectroscopy of $z=1-3$ ULIRGs from the Spitzer SWIRE
survey\thanks{Based on data obtained at the W. M. Keck Observatory, 
which is operated as a scientific partnership between the California 
Institute of Technology, the University of California, and NASA, 
and made possible by the generous financial support of the W. M. Keck Foundation.}}


\author{Stefano Berta\inst{1}\fnmsep\inst{2}\fnmsep\thanks{SB was supported by the
	Ing. Aldo Gini Foundation}
        \and
	Carol~J. Lonsdale\inst{2}\fnmsep\inst{3}
	\and
        Brian Siana\inst{4}
	\and 
	Duncan Farrah\inst{5}
	\and
	Harding~E. Smith\inst{2}
	\and
	Maria del~Carmen Polletta\inst{2}
	\and
	Alberto Franceschini\inst{1}
	\and
	Jacopo Fritz\inst{1}\fnmsep\inst{6}
	\and
	Ismael Perez-Fournon\inst{7}
	\and
	Micheal Rowan-Robinson\inst{8}
	\and
	David Shupe\inst{4}
	\and
	Jason Surace\inst{4}
}

\offprints{Stefano Berta, \email{berta@pd.astro.it, ste\_atreb@yahoo.it}}

\institute{Dipartimento di Astronomia, Universit\`a di Padova, Vicolo dell'Osservatorio 2, 
35122 Padova, Italy.
\and 
Center for Astrophysics and Space Sciences, University of California, 
San Diego, 9500 Gilman Dr., La Jolla, CA 92093-0424, USA.
\and
Infrared Processing \& Analysis Center, California Institute of Technology
100-22, Pasadena, CA 91125, USA. 
\and
Spitzer Science Center, California Institute for Technology, 220-6, Pasadena, CA 91125, USA.
\and
Astronomy Department, Cornell University, Itaca, NY 14853, USA.
100-22, Pasadena, CA 91125, USA. 
\and
INAF Osservatorio Astronomico di Bologna, Via Ranzani 1, 40127 Bologna, Italy.
\and
Instituto de Astrofisica de Canarias, 38200 La Laguna, Tenerife, Spain.
\and
Astrophysics Group, Blackett Laboratory, Imperial College London, Prince Consort Road, London SW7 2BW, UK.
}

\date{Received: November 22nd, 2006; accepted: March 8th, 2007}

\titlerunning{Keck spectroscopy}
\authorrunning{Berta S., et al. }

 
  \abstract
   {
   High-redshift ultra luminous infrared galaxies contribute the 
   bulk of the cosmic IR background and are the best candidates
   for very massive galaxies in formation at $z>1.5$.}
   {
   It is necessary to identify the energy source for their huge luminosities, 
   starburst or AGN activity, 
   in order to correctly interpret the role of ULIRGs in 
   galaxy evolution, and compute reliable estimates of their star formation rates, 
   stellar masses, and accretion luminosities.}
   {We present Keck/LRIS optical spectroscopy of 35 $z\ge1.4$ 
   luminous IR galaxies in the Spitzer Wide-area Infra-Red 
   Extragalactic survey (SWIRE) northern fields (Lockman Hole, ELAIS-N1, ELAIS-N2). 
   The primary targets belong to the ``IR-peak'' class of galaxies,
   having the 1.6$\mu$m (restframe) stellar feature detected in the IRAC
   Spitzer channels.
   The spectral energy distributions of the main targets are thoroughly analyzed, 
   by means of spectro-photometric synthesis and multi-component fits (stars + starburst
   dust + AGN torus). }
   {
   The IR-peak selection technique is 
   confirmed to successfully select objects above $z=1.4$, though some of the observed 
   sources lie at lower redshift than expected. 
   Among the 16 galaxies with spectroscopic redshift, 62\% host 
   an AGN component, two thirds being type-1 and one third type-2 objects.
   The selection, limited to $r^\prime<24.5$, is likely biased to optically-bright AGNs.
   All IR-peakers without emission lines have a non negligible continuum detection, 
   and are likely to be very
   powerful starbursts, heavily extinguished by dust ($A_V\ge5$ mag).
   The SEDs of non-AGN IR-peakers resemble those of 
   starbursts ($SFR=20-500$ M$_\odot$/yr) hosted in massive 
   ($M>10^{11}$ M$_\odot$) galaxies.
   The presence of an AGN component provides a plausible explanation for the 
   spectroscopic/photometric redshift discrepancies, as the torus 
   produces an apparent shift of the peak to longer wavelengths.
   These sources are analyzed in IRAC and optical-IR 
   color spaces. 
   In addition to the IR-peak galaxies, we present redshifts and 
   spectral properties for 150 objects, out of a total of 301 sources on slits.}
   {}


   \maketitle


\section{Introduction: distant starbursts in the SWIRE survey}

Locally rare, Ultra-luminous ($L_{IR} > 10^{12}$ L$_\odot$, ULIRGs)
and Hyper-luminous ($L_{IR} > 10^{13}$ L$_\odot$,
HLIRGs) infrared galaxies dominate the energy budget in the distant Universe 
\citep[e.g.][]{franceschini2001,elbaz2002}. 

Extragalactic surveys with ISO, SCUBA, MAMBO\footnote{Distant ULIRGs are commonly referred to as submillimeter
galaxies, or SMGs; here we reserve that term explicitly
for systems selected in submm or mm surveys, because distant ULIRGs
selected at other wavelengths may not necessarily
be (sub)mm-luminous.} (e.g. Franceschini et al., \citeyear{franceschini2003},
Genzel \& Cesarsky, \citeyear{genzel2000}, Ivison et al., \citeyear{ivison1998}; see Lonsdale et al., 
\citeyear{lonsdale2006}, for a review), and now {\em Spitzer}
have shown that the number of dusty, IR-luminous galaxies
at $2 < z < 3$ is several orders of magnitude higher than in the local Universe.
The analysis of the statistical properties of high-$z$ ULIRGs has shown 
that they contribute substantially to the cosmic infrared background (CIRB), 
discovered by COBE in the late '90s \citep{puget1996,hauser1998,elbaz2002,dole2006}.

The currently most successful models for galaxy formation
all invoke a ``biased'' hierarchical buildup
within a $\Lambda$CDM cosmology \citep[e.g.][]{cole2000,hatton2003,granato2004}
to assemble galaxies, suggesting that
the most massive objects (e.g. $M_{\textrm{stars}} > \textrm{ several }
10^{11}$ M$_\odot$) may assemble earlier, more quickly and
in richer environments than less massive  
ones. 
This may occur in short-lived, intense bursts of star formation at $z > 2$ 
 \citep{somerville2001,nagamine2005}.
This theoretical framework has enjoyed great success in describing
many observational results from the local Universe, and at moderate redshifts
\citep[e.g.][]{cole2005}. 

The sternest tests of these models, however, come from observations at $z>1$,
where the earliest formation stages of  massive galaxies and rich clusters are
predicted to occur. Several pieces of evidence exist that fully formed massive
galaxies were already in place at redshift $z>1.5$ 
\citep[e.g.][]{ellis1997,vandokkum2003}. 

Analyses of the ultra-luminous submillimeter galaxy (SMG) population at $z>1.5$
\citep[e.g.][]{chapman2004} support extreme star-formation rates in rare massive
objects. 
These distant IR sources, with implied $SFR>500$ M$_{\odot}/yr$ \citep[e.g.][]{farrah2002},
are the best candidates to be the progenitors of ellipticals which
formed most of their stars rapidly at $z\sim2-4$.  

Nevertheless, whether the huge luminosities of ULIRGs and SMGs are powered by starburst 
emission, AGN accretion or a combination of the
two has been often, and is still, a matter of debate. 
Discriminating between starburst and AGN power is fundamentally important 
for properly measuring their star formation rates, stellar masses, accretion rates, 
and also for understanding the connection between bulge and black hole
building. 

Spitzer \citep{werner2004} holds the key to this major question, because
the IR broadband spectral energy distributions (SEDs) and IRS \citep{houck2004} spectra 
of $z>1$ sources can discriminate warm 
AGN-dominated emission, characterized by a power-law
(torus-like) SEDs, from emission dominated by
stars \citep[e.g.][]{weedman2006}.   
\citet{lacy2004} and \citet{stern2005}
have demonstrated a strong segregation of AGN-dominated systems from
starburst-dominated galaxies in the IRAC color-color space.

The first major IRS surveys of the 24$\mu$m-brightest, optically-faint
galaxies from Spitzer surveys have shown
that they tend to be dominated by warm AGN-heated dust \citep{houck2005,yan2005}.
\citet{polletta2006} and Lonsdale et al. (in prep.) have used the 
added power of SED analysis to distinguish
the main energy source and estimate photometric redshifts for 
high redshift ULIRGs in the {\em Spitzer Wide-area InfraRed Extragalactic} Legacy survey
\citep[SWIRE][]{lonsdale2003,lonsdale2004}, and find that the 
AGN/starburst fraction decreases rapidly with decreasing 24$\mu$m 
flux, until starburst-dominated systems far exceed AGN-dominated ones 
as $f_{24}$ drops below 500$\mu$Jy (Lonsdale et al.).
\citet{weedman2006} have confirmed these
broadband AGN {\em vs.} starburst classifications, by exploiting 
IRS spectroscopy of 17 $z\sim2$ ULIRGs and finding that the 
broadband classifications are correct in 90\% of the cases.

Finally, \citet{farrah2006} have found evidence for significant clustering
in $z=2-3$ SWIRE starburst-dominated ULIRGs, suggesting that powerful star formation might  
indeed be taking place preferentially within high density environments.
Identifying powerful ULIRGs in the distant Universe is
particularly important, not only because they may be the most
massive galaxies in the process of formation, but also because they may
trace the rarest, most massive dark matter halos at $z\sim2$ ($M>
10^{13}$ M$_\odot$, density $< 10^{-7}$ Mpc$^{-3}$ at $z=2$). 
SWIRE has sufficient volume to include
about 85 halos with mass $> 10^{14}$ M$_\odot$ in the redshift range $z=2-3$ 
\citep{jenkins2001,mo2002}, which will evolve to
host extremely rich clusters of the Perseus class in the local Universe.
For comparison, all the Spitzer deep surveys (extended-GOODS, S-COSMOS, GTO Deep,
Dickinson et al. \citeyear{dickinson2003}, Sanders et al. \citeyear{sanders2007}, Fazio
et al. \citeyear{fazio2004b})
together potentially sample about 9 such haloes.
The First Look Survey (FLS, Soifer et al. \citeyear{soifer2004}) and GTO-Bootes survey 
\citep{eisenhardt2004} potentially sample 24 similar haloes, but they are 
shallower than SWIRE, thus the number of IRAC-selected detectable sources
is smaller. Finally, the sub-mm surveys to date have sampled 
a volume too small to identify even one halo with $M>10^{14}$ M$_\odot$.

We have therefore selected from SWIRE the brightest mid-IR examples of
starburst-dominated ULIRG candidates
in the $z=1.5-3$ range for spectroscopic observation at Keck, to confirm
their redshifts and characterize their nature.
Our selection of these systems is based on the detection within the
Spitzer IRAC bands of the redshifted stellar emission peak at $\sim$1.6$\mu$m in
galaxies \citep{sawicki2002,simpson1999}. We name these kind of sources ``IR-peakers'',
or ``IR-peak sources''.

We present here Keck/LRIS multi-object spectroscopy 
of 35 IR-peak systems in SWIRE northern fields. 
This instrument provides a nearly contiguous wavelength coverage from the 
3000 \AA\ atmospheric cutoff to the $Z$ band ($\lambda_c=9100$ \AA). 
The UV-optical throughput is larger than 50\% in the $U$ to $g$ bands. 
This represents a great advantage over optical-only spectroscopic surveys:
Ly$\alpha$ can be probed to $z\ge1.7$ and {\sc [Oii]} ($\lambda=3727$ \AA) can be detected at $z\le1.4$. 
Basically, only a very narrow redshift desert exists for IR-peak galaxies  observed with LRIS.
Moreover, on the 10m Keck telescope, 
LRIS allows the detection of $r^\prime\simeq24$ galaxies in reasonably fast 
exposure times ($<$ 2h).

Several other categories of interesting SWIRE targets have been included 
in the observations, in order to fill
slitlet masks: X-ray, radio, power-law, extremely red, 
and 24$\mu$m sources.

Section {\bf 2} describes the sample selection and the available data in the SWIRE 
fields; Sect. {\bf 3}  deals with observations and data reduction.
Spectroscopic results are presented in Sect. {\bf 4}, where we also discuss
photometric redshifts. Details on the IR-peak population are given
in Sect. {\bf 5}, including spectroscopy and modeling of broad-band SEDs.
Section {\bf 6} includes the description of some additional interesting objects:
high redshift quasars and X-ray sources.  
Finally, Sects. {\bf 7} and {\bf 8} discuss our findings and summarize our conclusions.
Throughout this paper we adopt a $H_0=71$ $[$km s$^{-1}$ Mpc$^{-1}]$, 
$\Omega_m=0.27$, $\Omega_\Lambda=0.73$ cosmology.

\section{Sample Selection}

The SWIRE Legacy survey \citep{lonsdale2003,lonsdale2004,surace2004}
covers 49 deg$^2$ in all seven
Spitzer imaging bands, and has detected over 2 million 
galaxies up to $z>3$. 
This sensitivity means that SWIRE can detect tens of thousands of
star-forming galaxies with SFR $\sim$ hundreds M$_\odot$/yr, such as typically
found in
blank-field submm surveys, in at least two Spitzer bands at $z\sim1-3$.
Such a large ULIRG sample will include not only systems similar to the
sub-millimeter galaxy class (SMGs),  but also
objects dominated by significantly warmer dust than typical of the
submm- and mm-selected systems.

\subsection{Spitzer data}

The SWIRE northern fields benefit from extensive multiwavelength coverage,
over the whole electromagnetic spectrum from the X-rays to radio
frequencies. The SWIRE datasets are widely described by \citet{surace2004}.

The SWIRE Lockman Hole field is centered at RA$=$10h45m00s,
DEC$=$+58d00m00s, with a total area of 10.6 deg$^2$. 
Observations with the Infrared Array Camera \citep[IRAC,][]{fazio2004}
were obtained in April 2004, 
and the Multiband Imaging Photometer \citep[MIPS,][]{rieke2004}
data were collected in May 2004.

The SWIRE ELAIS-N1 field is centered at RA$=$16h11m00s, 
DEC$=$+55d00m00s, and covers 9 deg$^2$; Spitzer observations were 
carried out during January and February 2004.

The SWIRE ELAIS-N2 field is centered at RA$=$16h36m48s, 
DEC$=$+41d01m45s, over 4 deg$^2$ and was observed in July 2004.

Data processing is described by \citet{surace2004} and 
Shupe et al. (in prep.), and consists of Basic Calibrated Data (BCD) by the SSC
pipeline plus post-processing aimed at artifact removal, mosaicking and
source extraction.
The mosaicking was performed with the SSC routine MOPEX, and
source extraction with SExtractor \citep{bertin1996}.  
IRAC fluxes were extracted through a 1.9${\arcsec}$ diameter aperture and corrected 
to total fluxes following SSC prescriptions; MIPS fluxes 
were extracted by means of PRF fitting (see Surace et al., and MIPS Data Handbook 2006).
The 5$\sigma$ depths (consistent with the 90\% completeness levels)
of the Spitzer data are on average 3.7, 7.4, 43, 46, and
195 $\mu$Jy at 3.6, 4.5, 5.8, 8.0, and 24 $\mu$m, respectively 
(Surace et al., in prep.), with field to field variations.
None of the primary targets is detected at 70 and 160 $\mu$m in the SWIRE survey, 
at the average 5$\sigma$ flux limits of 17.5 and 112 mJy respectively.

\subsection{Available optical data}

The Lockman Hole field was observed in the 
$U$, $g^\prime$, $r^\prime$, and $i^\prime$ bands with
the MOSAIC
Camera at the Kitt Peak National Observatory (KPNO) Mayall
4m Telescope, February 2002 ($g^\prime$, $r^\prime$, and $i^\prime$) 
and January 2004 ($U$ band). The scale of the Camera is 0.26${\arcsec}/ \textrm{pix}$
and the field of view is $36^\prime\times36^\prime$. The astrometric mapping 
of the optical MOSAIC data is good to less than 0.4${\arcsec}$ and the seeing varied 
between 0.9 and 1.4 arcsec.
Data reduction was performed with the  Cambridge Astronomical Survey Unit
\citep[CASU,][]{irwin2001} pipeline, following the procedures described
in \citet{babbedge2004}. 
Fluxes were measured within a 3${\arcsec}$ aperture
(diameter) and corrected to total fluxes using growth curves.
Typical $5\sigma$ magnitude 
limits are 24.1, 25.1, 24.4 and 23.7 in $U$, $g^\prime$, $r^\prime$ and $i^\prime$ 
respectively (Vega), for point-like sources. 

The ELAIS-N1 and EN2 fields were observed in the $U$, $g^\prime$, $r^\prime$, $i^\prime$ and $Z$ bands, as
part of the 2.5m Isaac Newton Telescope (INT, Roque de Los Muchachos, 
La Palma, Spain) Wide Field Survey \citep[WFS,][]{mcmahon2001}.
The data were processed with the
CASU pipeline; the average limiting magnitudes (Vega, 5$\sigma$) 
across the fields are 23.40 ($U$), 24.94 ($g^\prime$), 24.04 ($r^\prime$), 23.18 ($i^\prime$) and
21.90 ($Z$).
The overall photometric accuracy of the INT WFS survey is 2\%
Further details are given in \citet{babbedge2004} and \citet{surace2004}.

\subsection{X-ray and Radio data}

A 0.6 deg$^2$ sub-area of the Lockman hole field, centered
at RA$=$10h46m, DEC$=$59d01m was observed with the 
Chandra Advanced CCD Imaging Spectrometer \citep[ACIS-I,][]{weisskopf1996}
in the X-rays, during September 2004. Description of 
observations and data analysis is provided in \citet{polletta2006}.
The total exposure time was 70 ks, reaching 
3$\sigma$ fluxes of $\sim 10^{-15}$, $5\times 10^{-16}$, and $10^{-14}$
$[$erg cm$^{-2}$ s$^{-1}]$ in the broad (0.3$-$8 keV), soft (0.3$-$2.5 keV) 
and hard (2.5$-$8 keV) bands respectively. 

As part of the ELAIS Deep  X-ray Survey (EDXS), 
a sub-region of ELAIS-N1 was targeted by the Chandra ACIS instrument.
Observations and data analysis are described in \citet{manners2003,manners2004}
and \citet{franceschini2005}.
The Chandra field is centered at RA$=$16h10m20.11s, DEC$=$+54d33m22.3s (J2000.0) 
and the total net exposure time is 71.5 ks (after flare cleaning).
Sources were detected to flux levels of $2.3\times10^{-15}$,  
$9.4\times10^{-16}$ and $5.2\times10^{-15}$ $[$erg cm$^{-2}$ s$^{-1}]$
in the 0.5$-$8 keV, 0.5$-$2 keV and 2$-$8 keV bands.

Finally, a deep, 1.4 GHz radio survey, centered at RA$=$10h46m, DEC$=$59d01m
covers $40^\prime\times40^\prime$ in the Chandra/SWIRE Lockman Hole field.
These data were obtained in multiple Very Large Array (VLA) runs, 
obtained in Dec. 2001, Jan.-Mar. 2002, and Jan. 2003 (Owen et al., in prep.)
with configurations A/B/C and D. The total integration time spent
on source is 500 ks. The r.m.s. noise at the image center
is 2.7 $\mu$Jy \citep[see also][]{polletta2006}.

\subsection{Primary targets}

Primary targets were selected in order to include SWIRE $z>1.5$ ULIRG
candidates. 
The near-IR restframe spectral energy distribution (SED) of galaxies is
characterized by  
a peak at 1.6 $\mu$m, due to the Planck spectrum of low-mass stars (dominated
by M-type), enhanced by a minimum in the H$^-$ opacity in
stellar atmospheres \citep{sawicki2002}. On the red side of the peak, molecular absorption bands 
further blue the (H--K) color.
This peak is fully characterized
by the IRAC instrument if at least one of the IRAC photometric bands (3.6,
4.5, 5.8, 8.0 $\mu$m) falls long- or shortward of the peak. This happens
when the peak lies in the 4.5 or 5.8 $\mu$m band, i.e. for redshifts
in the range $z=1.4-3.0$.

IRAC was in part designed for photometric selection of galaxies at these
redshifts displaying this feature \citep{simpson1999}. 
\citet{egami2004} have shown that starburst-dominated SMGs show this stellar
population features strongly in the IRAC SEDs. 

We have selected ``IR-peak'' sources by exploiting SWIRE Spitzer
photometry, isolating objects with SEDs peaking at 4.5 or 5.8 $\mu$m, in the SWIRE
Lockman Hole, ELAIS-N1 and N2 fields.
The density on the sky of these sources is about 200 deg$^{-2}$, at 
the SWIRE flux limits (Lonsdale et al., Berta et al., in prep.).

All sources are detected in the 3.6, 4.5 and 5.8 $\mu$m bands; 
for some only an upper limit to the 8.0 $\mu$m flux is available 
(see Tab. \ref{tab:IB_phot}).
In the latter case, this upper limit 
is required to be consistent with the IR-peak definition, i.e. lower than
the 5.8$\mu$m measured flux. 

\begin{figure*}[!ht]
\centering
\includegraphics[width=0.43\textwidth]{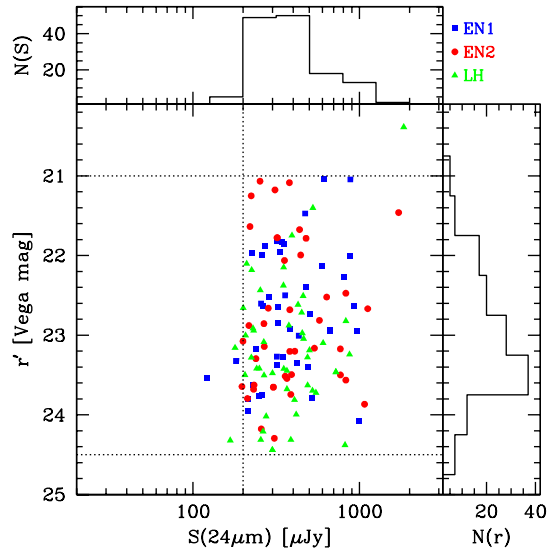}
\caption{Selection of sources for our spectroscopic Keck observations:
distribution of observed targets in the 24$\mu$m {\em vs} $r^\prime$
space.}
\label{fig:selection}
\end{figure*}

Optical magnitudes were limited to the $r^\prime<24.5$ (Vega) range, in order to
include sources bright enough to be detected with LRIS. Moreover objects
brighter than $r^\prime<21$ were avoided, in order to minimize the contamination
by low redshift foreground sources (see Fig. \ref{fig:selection}).

On average, 4 IR-peak galaxies were put onto a slit per LRIS
mask (see Tab. \ref{tab:obs}).  Table \ref{tab:IB_phot} reports the 
basic data for the selected IR-peak targets.The total number of IR-peakers 
observed is 35.

\subsection{Mask fillers}\label{sect:fillers}

LRIS masks can host as much as 30 slitlets, 
the exact number depending on the positions of the selected targets on the sky. 

In addition to the primary IR-peakers included in each mask, 
the remaining slitlets were filled with sources from the SWIRE
catalogs showing interesting photometric
multi-wavelength properties, such as red
optical-NIR colors ($[i-4.5]\ge5$), consistent with $z\sim1$ systems,
X-ray or radio detection, IRAC red power-law (AGN-like) with a
monotonic slope, and finally generic 24 $\mu$m detection.

Column eight in Tabs. \ref{tab:redshifts1} and \ref{tab:redshifts3}
reports a rough classification 
of the most interesting sources, based on their photometric properties.
In particular, we distinguish: 4.5$\mu$m- and 5.8$\mu$m-peak galaxies 
(P2, P3, P3L\footnote{P3L sources have IRAC 3.6, 4.5, 5.8 $\mu$m detection, but only
an upper limit at 8.0 $\mu$m.}),
X-ray and radio sources (X and R in column 8), and objects  with a monotonic 
power-law like IRAC SED (pow). 

The total number of mask is 4 in the Lockman Hole field,
which have been observed during the first three hours of each night, three
in ELAIS-N1 and three in ELAIS-N2. A total of 235 slits were defined.
Figure \ref{fig:selection} shows all targets detected 
in the mid-IR, in the $r^\prime$-band {\em vs.} 24$\mu$m space.

\begin{figure*}[!ht]
\centering
\includegraphics[width=0.43\textwidth]{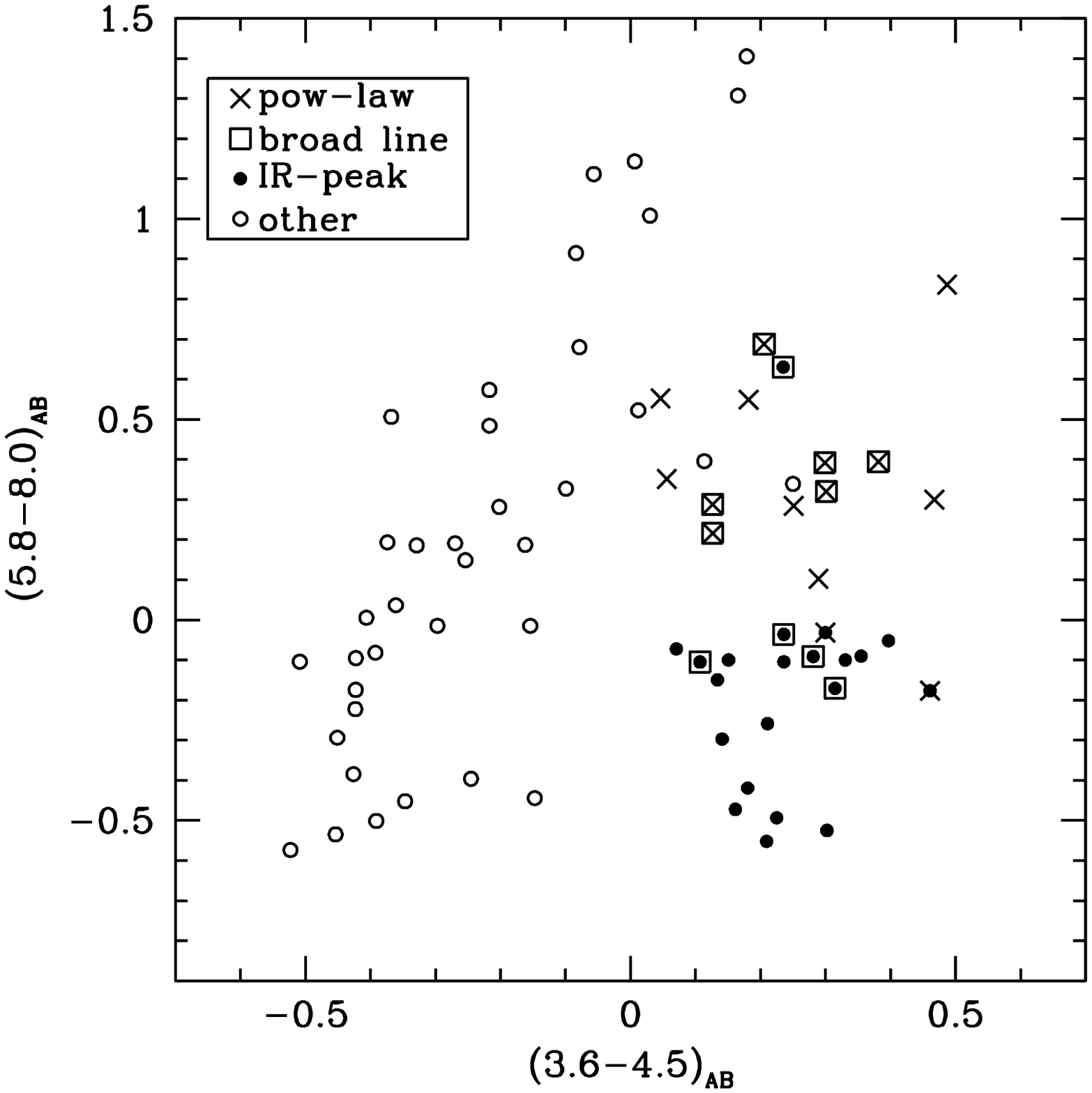}
\includegraphics[width=0.43\textwidth]{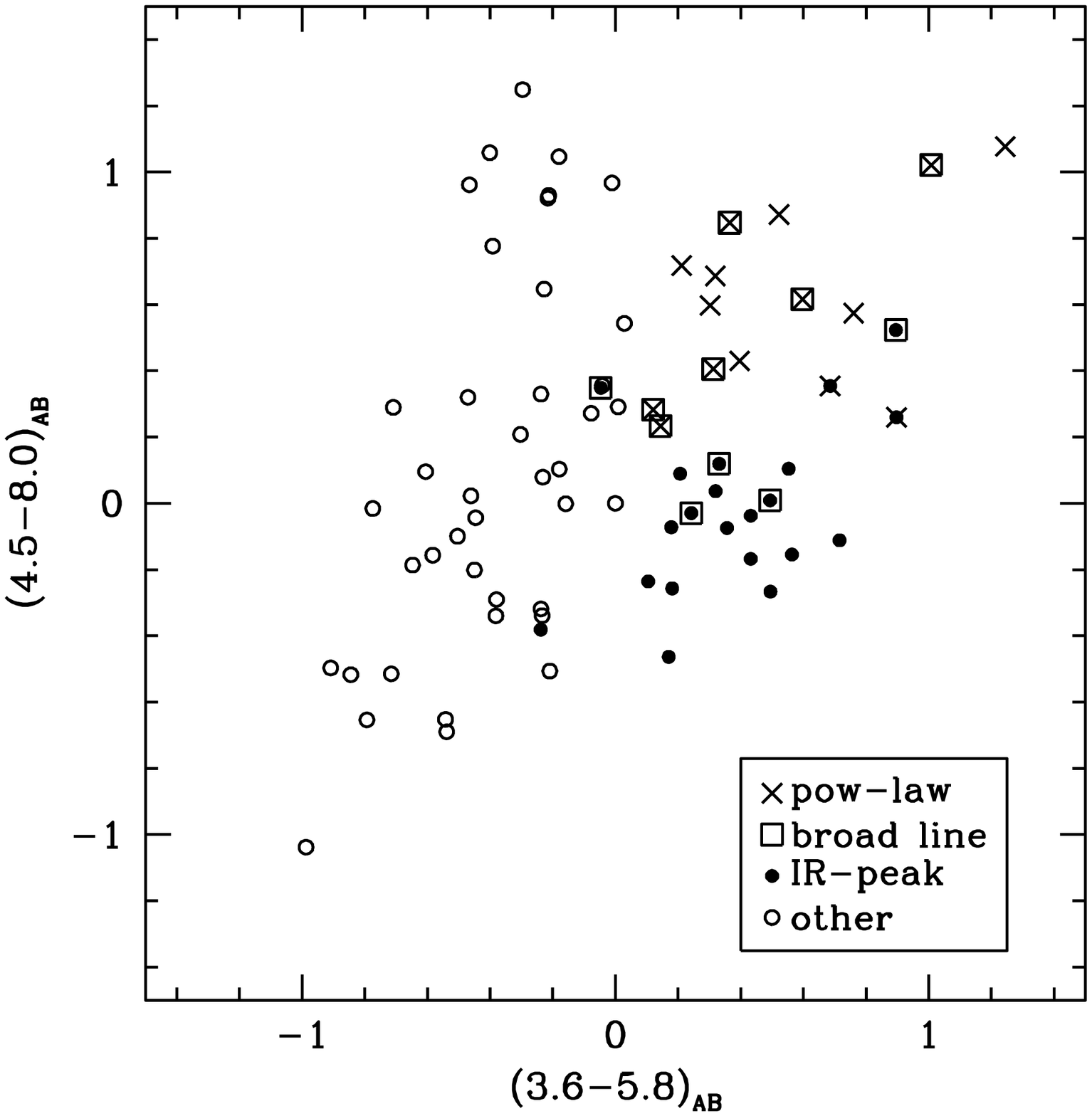}
\caption{Distribution of the observed targets in IRAC color space \citep{stern2005,lacy2004}. 
We distinguish: IR-peak sources (filled circles),  
power-law targets (crosses), broad-line detections (open squares). All other cases 
are plotted as open circles.}
\label{fig:ib_r_f24_lacy}
\end{figure*}

These masks include 35 IR-peak targets (9 4.5$\mu$m and 26 5.8$\mu$m peakers),
7 X-ray sources, 12 radio sources, 19 IRAC power-law objects and 139 objects 
detected in the MIPS 24$\mu$m channel.
Figure \ref{fig:ib_r_f24_lacy} shows the distribution of the selected targets in the 
IRAC color-space \citep{lacy2004,stern2005}; different symbols refer 
to different photometric (and spectroscopic) properties: IR-peak sources
(filled circles), power-law IRAC SEDs (crosses), broad-line objects (open squares).
It is already worth noting that the power-law and broad-line classifications 
are $\sim100$\% consistent with each other, these targets lying in 
the locus of AGNs in the IRAC color space (Lacy et al., See also Sect. 
\ref{sect:IR-peakers}).

\section{Observations and data reduction}

Observations were carried out in multi-object mode with the Low Resolution Imaging Spectrometer
\citep[LRIS,][]{oke1995} at the Cassegrain focus of the Keck-I
telescope, during the nights of May 27th and 28th, 2006.

The LRIS instrument makes use of a dichroic to split the incoming light into 
a blue and red beam. Gratings, grisms and filters can be changed independently
for the two beams. We have adopted the dichroic designed to split light at
5600 \AA. 

As far as the blue arm is concerned, in order to obtain a maximum throughput
in the spectral range 3200-5000\AA, we used the 400 lines/mm grism, blazed at 3400 \AA,
providing a good throughput from the atmospheric cutoff at 3000 \AA\ to the dichroic 
5600 \AA\ limit.

On the red arm, we used the  400/8500 grating, blazed at
7400 \AA, providing wavelength coverage up to $\sim 9550$ \AA.
This configuration was chosen  in order to optimize spectral coverage of
simultaneous LRIS red and blue observations, as well as wavelength calibration
with Hg, Cd, Zn, Ar arc lamps. The effective spectral coverage depends on the
positioning of slitlets in the mask, relative to the telescope focal plane.

The dispersion in the blue and red arms is 1.09 \AA/pix and 1.86 \AA/pix. 
A 1.2 arcsec slit was adopted,  
resulting in an instrumental resolution (measured as the FWHM of arc lines)
of $\sim10.5$ \AA. This corresponds to 750 and 420 $[$km s$^{-1}]$ at 
4200 and 7500 \AA, respectively. The seeing varied between $\sim1.0$ and $\sim$1.3 during the
two observing nights.

The slit masks cover an effective area of $6\times8$ arcmin$^2$ on the sky;
between 15 and 30 slitlets with variable length were placed per mask, this number
depending on the sky distribution of the selected targets.
A total of 10 masks was observed during the two night run, with 
exposure times between 3600s and 5400s, split into three exposures per mask.
Table \ref{tab:obs} lists the position of the pointings on the sky, as well as exposure times,
number of slitlets and number of IR-peak targets included.

Spectro-photometric standard stars Feige34 and BD+28D4211 were observed during
the nights, taking care to have as close an airmass as possible to the science
pointings. Flat field and arc lamp frames were taken at the same telescope
position (ALT,AZ) as the science spectra, in order to reproduce the same
instrumental flexures and shifts and avoid troublesome corrections during data
reduction. 
Arc-frames were obtained using Hg, Cd, Zn,
Ar lamps, ensuring bright calibration lines over the whole spectral range from
3000 \AA\ to 9500 \AA, with a gap between 5500 and 6500 \AA\ only. 


Data reduction was carried out by using the standard tasks in the
IRAF\footnote{The
package IRAF is distributed by the National Optical Astronomy Observatory which
is operated by the Association of Universities for Research in Astronomy, Inc.,
under cooperative agreement with the National Science Foundation.} environment.
Bias, dark and flat field corrections ware performed in the standard manner, by
using the overscan CCD regions and the dome flat field frames obtained at the
telescope, as well as the gain values reported on the LRIS
webpage\footnote{http://www2.keck.hawaii.edu/inst/lris/}
for the different amplifiers.
Wavelength and flux calibration were performed on the two-dimensional spectra, 
after background subtraction. 

Extraction of spectra was performed in all cases where a continuum trace was
detected; lines were identified on non-flux-calibrated frames, in order to avoid
losses of spectral coverage due to the relative position of slits with respect
to the standard star spectrum. Line properties were measured after flux calibration.

The lines detected for the IR-peak galaxies are listed in Tabs.
\ref{tab:spec_IB_1} and \ref{tab:spec_IB_2}, where we list observed
wavelengths, equivalent widths, FWHMs (corrected in quadrature for the instrumental 
resolution) and derived redshifts.

Spectroscopic redshifts of all the observed targets are listed in Tabs.
\ref{tab:redshifts1} and \ref{tab:redshifts3}, where we include also the number
of emission/absorption lines detected for each object.

\begin{table*}[!ht]
\centering
\begin{tabular}{l c c c c c c c c}
\hline
\hline
Mask & RA & DEC & P.A. & A.M. & t(exp) & N & N & N\\
 & J2006.4 & J2006.4 & $[$deg$]$ & & $[$s$]$ & (slits) & (peak) & (reds$^\ast$) \\
\hline
LH\_247597 & 10:52:32.694  & +57:46:27.896 & 110 & 1.55 & 3600 & 21 & 5 & 6 \\	
LH\_575325 & 10:47:43.555  & +59:09:22.974 & 131 & 1.35 & 4500 & 26 & 5 & 15 \\	
EN1\_205467 & 16:09:42.168 & +54:39:49.085 & 155 & 1.25 & 3600 & 24 & 3 & 18 \\	
EN1\_282078 & 16:16:56.673 & +55:32:02.568 & 179 & 1.26 & 5400 & 24 & 2 & 18 \\	
EN2\_273717 & 16:33:33.569 & +40:54:40.722 & 143 & 1.21 & 5400 & 23 & 3 & 15 \\	
\hline
LH\_128777 & 10:59:03.752  & +57:46:56.282 & 105 & 1.56 & 3600 & 29 & 3 & 11 \\	
LH\_579894 & 10:48:09.714  & +59:09:25.529 & 119 & 1.36 & 4500 & 24 & 6 & 9 \\	
EN1\_341469 & 16:03:38.599 & +54:23:52.565 & 180 & 1.23 & 5400 & 24 & 4 & 16 \\	
EN2\_10334 & 16:42:09.003  & +40:45:51.746 & 104 & 1.22 & 5400 & 25 & 2 & 20 \\	
EN2\_172324 & 16:34:50.206 & +41:00:10.745 & 145 & 1.08 & 4500 & 15 & 4 & 11 \\	
\hline
\multicolumn{9}{l}{$^\ast$: without accounting for serendipitous sources}\\
\end{tabular}
\caption{Summary of observations: each mask is named with the identification
number of its primary target. The number of slitlets, of IR-peakers included, and  
of measured spectroscopic redshifts are reported.}
\label{tab:obs}
\end{table*}


\section{Results}

The full sample of targeted sources includes 233 objects\footnote{Note that a
couple of targets were observed twice, in distinct masks, therefore the
effective number of targets is 233 in 235 slitlets.}, distributed
in the Lockman Hole, ELAIS-N1 and ELAIS-N2 SWIRE fields.

We have computed redshifts on the basis of the presence of 
emission lines, both in the ultraviolet and optical restframe domains, such as
Lyman-$\alpha$ ($\lambda=1216$ \AA), {\sc Niv} ($\lambda=1240$ \AA), {\sc Oi}
($\lambda=1304$ \AA), Si{\sc
iv, Oiv}$]$ ($\lambda=1400$ \AA), {\sc Niv}$]$ ($\lambda=1486$ \AA), {\sc Civ}
($\lambda\lambda=1548,1551$ \AA),  He{\sc ii} ($\lambda=1640$ \AA), 
{\sc Ciii}$]$ ($\lambda=1909$ \AA), Mg{\sc ii} ($\lambda\lambda=2796,2803$ \AA), Fe{\sc
ii} and Fe{\sc iii} lines ($\lambda=2000-3000$ \AA), $[${\sc Oii}$]$
($\lambda\lambda=3726,3729$ \AA),  
as well as Balmer Hydrogen emission and absorption, Ca{\sc ii-HK}, $[${\sc
Oiii}$]$, $[${\sc Nii}$]$, and $[${\sc Sii}$]$ lines for lower redshift sources.
We do not have adequate resolution to resolve the $[${\sc Oii}$]$, {\sc Civ} and Mg{\sc ii} doublets.

The spectroscopic success rate per mask strongly depends on 
the observing conditions, such as presence of cirrus, seeing and airmass.
The last column in Tab. \ref{tab:obs} lists the number of redshifts
obtained for each mask, without accounting for serendipitous sources.

Lockman Hole masks were observed during the first hours of each night, at
increasing airmass, with relatively poor results (see Tab. \ref{tab:obs}).
During the first night, 6/21 and 15/26 sources have a successful redshift
estimate for masks LH\_247597 and LH\_575325 respectively\footnote{These numbers don't
include serendipitous sources}. 
Masks LH\_128777 and LH\_579894, observed on the second night, turn out to
have 11/29 and 9/24 good spectroscopic redshifts, without taking into account
serendipitous sources.

It is worth to note that mask LH\_579894 has a success rate as low as 
LH\_128777 ($\sim38$\%), despite the lower airmass (1.36 {\em vs.} 1.56). 
The main reason for this effect is that the former contains fewer 24$\mu$m-bright
targets than the latter (45\% {\em vs.} 60\% of the objects in slit), resulting 
in a lower emission lines detection rate.

The ELAIS fields were observed at lower airmasses, hence the success rate 
is higher for these areas, being 73\% on average, with a 80\% peak in the best
case.
A total of 139 redshifts have been derived for the targeted objects.

The redshift uncertainty depends on the number of detected spectral features.
In the case that only one emission line is detected, the sources of uncertainty
on the redshift estimate are given by the wavelength calibration of the spectrum
(having a typical r.m.s of $0.7-1.0$ \AA) and the centroid uncertainty in 
positioning during the gaussian fit to the line profile (which is a fraction of a pixel 
and negligible with respect to estimating redshifts). When dealing with multiple line 
detections, the average $z$ is computed and the uncertainty is given by the 
dispersion of the average. Typical uncertainties on $z$ are thus 
smaller than $\Delta z=0.01$ for narrow lines. In the case of broad emission lines, 
a lorentzian fit to the lines was usually adopted, but the asymmetry and broadness of profiles
cause the uncertainty of line positioning to be larger and dominate the $\Delta z$.
In this case, the redshift uncertainty can be as large as $\Delta z=0.03$.

In addition to the formally targeted objects, 68
serendipitous sources were detected, and for 35 of these a redshift estimation
was possible. We have identified these serendipitous objects by measuring their
projected distance from the main target in the same slit, matching it to the
SWIRE multi-wavelength catalog, and visually seeking for SWIRE counterparts on
$r^\prime$ band and 3.6$\mu$m images. Only 15 were identified, while another 14
are detected in the optical but not by Spitzer. Among these, only 11 have 
a spectroscopic redshift.
In Tabs. \ref{tab:redshifts1} and \ref{tab:redshifts3} we list only those 
serendipitous sources with a SWIRE/Spitzer counterpart.

The total number of redshifts available is 174, for a total of 301 objects in slits;
150 sources with redshift have a SWIRE identification.

\subsection{Photometric redshifts}

Taking advantage of the extensive multiwavelength coverage available in the 
observed areas, we have computed photometric redshifts for all the 
targeted sources, by using the Hyper-z \citep{bolzonella2000} and the 
\citet{rowanrobinson2003} codes.

In the former case, we have adopted a semi-empirical template library including GRASIL \citep{silva1998}
models of spiral and elliptical galaxies,
M82 and Arp220 templates (Silva et al.) upgraded with observed PAH mid-IR features, 
a ULIRG template \citep[IRAS 19254-7245,][]{berta2003}, type-1 AGN and Seyfert templates by Polletta et al. (in prep.),
obtained by averaging observed AGN SEDs.
The fits were performed using 
only the optical and IRAC ($3.6-8$ $\mu$m) data, ignoring the MIPS 24$\mu$m
flux. Tests including the 24$\mu$m data have been attempted as well, but they have shown a 
higher degree of degeneracy and aliases in the photometric redshift estimate.

\begin{figure*}[!ht]
\centering
\includegraphics[width=0.9\textwidth]{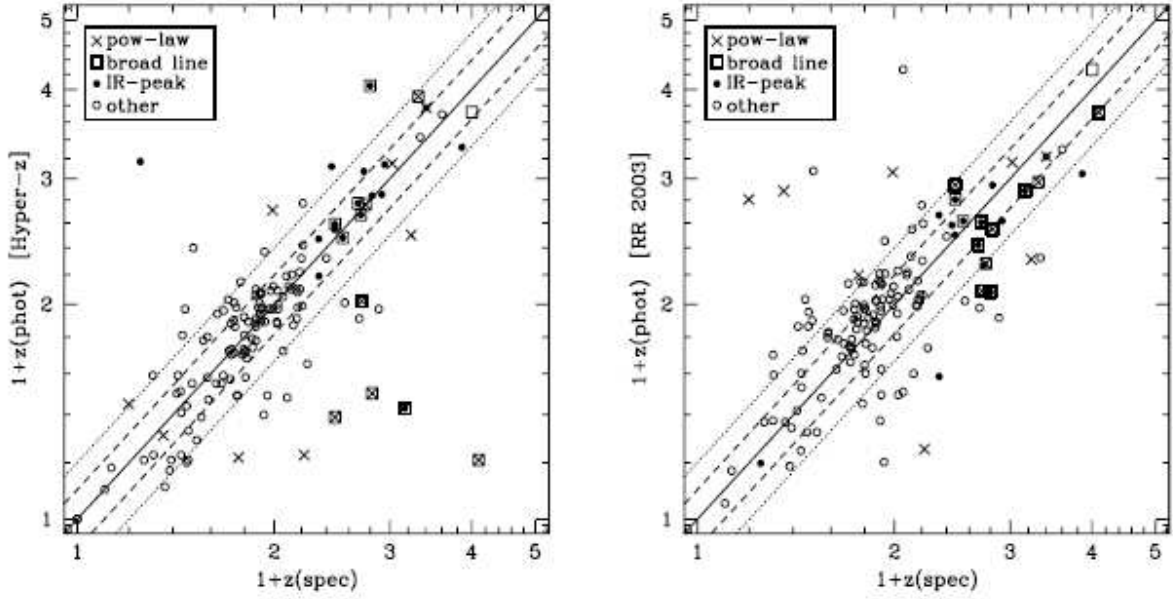}
\caption{Comparison between spectroscopic and photometric redshifts.
Dashed and dotted lines represent $\pm10$ and 20\% uncertainties. 
{\em Left panel}: results obtained with Hyper-z \citep{bolzonella2000}.
{\em Right panel}: results of the \citet{rowanrobinson2003} code.}
\label{fig:z_spec_phot_2}
\end{figure*}

The photometric redshifts thus obtained are compared to the spectroscopic Keck 
results in Fig. \ref{fig:z_spec_phot_2} (left panel). The dashed and dotted lines represent 10\% and 20\% 
uncertainty, respectively. Filled circles represent IR-peak sources, crosses are objects 
with power-law IRAC SEDs, and open squares indicate broad-line detections. All other cases 
are plotted as open circles.

Outliers are mostly AGNs (crosses or open squares in Fig. \ref{fig:z_spec_phot_2}), 
typically showing a power-law spectral energy distribution 
from the optical to the mid-IR. In this kind of object, neither strong, nor sharp features are
detected in the broad band SEDs, therefore the photometric redshift estimate 
often fails. 

Accounting for all sources with a spectroscopic redshift, the r.m.s. of the distribution\footnote{Here 
$\Delta$ is the difference between the photometric and spectroscopic redshift estimates.}
of $\Delta\left(1+z\right)$ is 0.095. Excluding power-law sources, it decreases to 0.069.
The semi inter-quartile range (s.i.q.r.) computed for all sources is 0.028.

The results obtained by using the \citet{rowanrobinson2003} code are shown in 
the right panel of Fig. \ref{fig:z_spec_phot_2}. 
As far as AGN-dominated objects are concerned, the results of this code show
a much better consistency between photometric and spectroscopic redshifts
for high-redshift sources, while it seems to fail for low-redshift
ones. For the latter, the photometric redshift is overestimated. 
The adopted templates are those described in \citet{rowanrobinson2004},
including AGN SEDs built on actual ELAIS data. The overall concordance of the photometric 
estimate and the spectroscopic measure of redshifts is similar to the one obtained
with Hyper-z, having a $\Delta(1+z)$ r.m.s. and s.i.q.r. of 0.091 and 0.043.
The number of dramatic failures (outliers) is smaller than in the Hyper-z 
case, but the median scatter is slightly larger.

\begin{figure*}[!ht]
\centering
\includegraphics[width=0.43\textwidth]{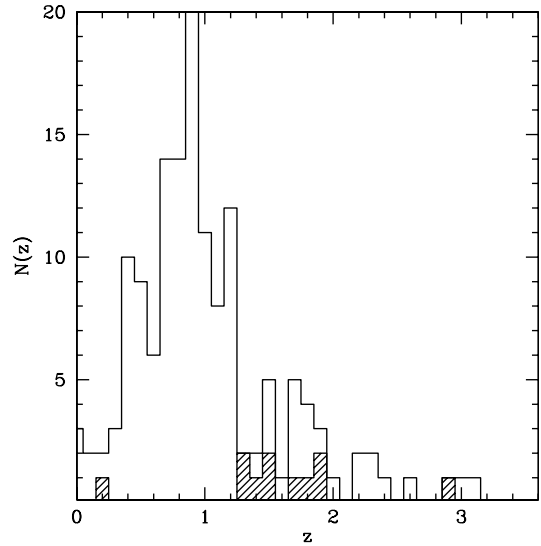}	
\caption{Redshift distribution of successfully detected sources (spectroscopic $z$).
The empty histogram shows spectroscopic redshifts for all sources, while
the shaded histogram represents IR-peak sources alone.}
\label{fig:zdistr}
\end{figure*}

Finally, Fig. \ref{fig:zdistr} reports the redshift distribution of our targets with 
spectroscopic redshift. 
The white histogram represents the distribution of spectroscopic redshifts 
for all sources, while the shaded histogram includes 
only IR-peak objects.

\subsection{Spectroscopic classification}\label{sect:spec_class}

The last column in Tables \ref{tab:redshifts1} and \ref{tab:redshifts3} 
reports the spectral classification of targets, based on the detected lines.
We classify as simply ``emission line'' galaxies (ELG), those sources with 
insufficient 
lines to apply any diagnostic technique, i.e. sources with only one or two
lines detected. Similarly, ``absorption line'' galaxies (ALG) have only absorption lines detected.

When possible, emission line fluxes were corrected for extinction, as
derived from the observed Balmer decrement (using the available 
Balmer lines and assuming case-B recombination, Hummer \& Storey, \citeyear{hummer1987}).

We classify starburst galaxies (SB) on the basis of different criteria:
\begin{itemize}
\item presence of strong emission lines (e.g. {\sc [Oii]}, $\lambda=3727$ \AA) and 
young star absorption lines (e.g. type-A stars), such as the advanced Balmer series, 
from H$\delta$ down to H-10;
\item conformity to optical diagnostic diagrams 
for AGN/starbursts, based on optical emission lines 
\citep{veilleux1987,dessauges2000,baldwin1981};
\end{itemize}

Type-1 AGNs (BLAGN flag) are recognized through the presence of broad emission lines
(FWHM$>1000$ $[$km s$^{-1}]$), both in the ultraviolet and optical
restframe spectral domains.

Type-2 AGNs (NLAGN) are identified 
by the presence of high-ionization narrow emission lines 
(e.g {\sc Nv}, {\sc Civ}, He{\sc ii}, $[$Ne{\sc v}$]$, $\lambda=1240,\ 1549,\ 1640,\ 3426$ \AA) 
in the observed spectra \citep[e.g.][]{farrah2005,villar1996,allen1998}, 
or on the basis of optical diagnostic diagrams
\citep[e.g.][]{veilleux1987,baldwin1981} in some cases. 
Because only a few optical emission lines are available, a distinction between 
Seyfert-2 galaxies and LINERs is not possible, apart in one case (EN1\_340460, 
at the boundary between starbursts and LINERs).
It is worth specifying that ultraviolet emission lines
such as {\sc Civ} can be produced also by star forming activity, typically
being heated by O stars, but are always associated with a P-Cyg profile, produced 
by stellar winds with velocities higher than 1000 $[$km s$^{-1}]$ 
\citep[e.g.][]{shapley2003,farrah2005}.
Given the resolution of our spectra, when no P-Cyg profile is detected and 
the {\sc Civ} ($\lambda 1549$ \AA) line is observed as narrow emission only,
a type-2 AGN explanation is favored.

Globally, 122 narrow-line emission galaxies have been 
classified. Among these, 44 have enough spectroscopic information to allow
a unambiguous classification: 39 turn out to be starbursts and 5 type-2 AGNs,
at least in the sampled optical range. Seven galaxies show only absorption lines (typically
the Ca{\sc ii}-HK doublet and some advanced Balmer Hydrogen lines), and 17 are type-1 AGNs. Finally, four stars were 
identified in our slitlets.

\section{IR-peak galaxies}\label{sect:IR-peakers}

As far as IR-peak sources are concerned, a total of 35 targets were
included in our 10 masks; among these, 11 galaxies have a confirmed
spectroscopic redshift in the range $z=1.5-3.0$, 5 
lie between $1.0<z<1.5$, and one turned
out to be a low redshift ($z<1.0$) (confused, see below) interloper. For the remaining 
18 IR-peak targets no spectral features were
detected, but in all cases a non negligible continuum is present (columns 2-3 in Tabs.
\ref{tab:spec_IB_1} and \ref{tab:spec_IB_2} specify
if a continuum was detected in the blue and/or red LRIS arms and over which wavelength range). 

\begin{landscape}

\begin{figure}[!ht]
\centering
\rotatebox{-90}{\includegraphics[width=0.90\textwidth]{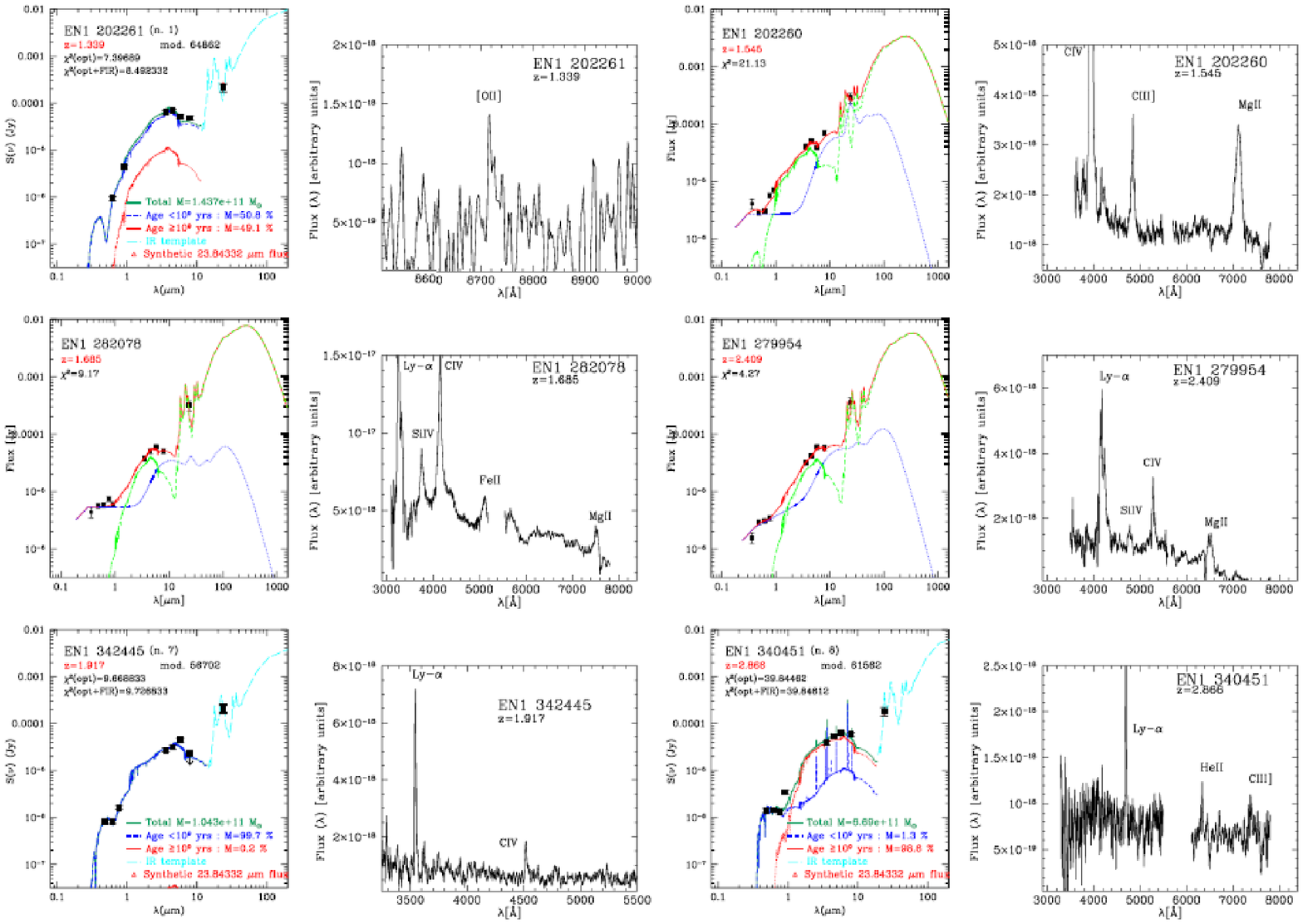}}
\caption{IR-peak sources with spectroscopic redshift: broad band SEDs and spectra.
Sources with no AGN detected in the UV-optical spectra are fitted with 
spectrophotometric synthesis \citep{berta2004}. In these cases, the dotted lines
represent old (age $\ge10^9$ yr) stellar populations, dashed lines are younger stars,
the dashed lines long ward 5$\mu$m (restframe) are the IR starburst template
adopted (see text) and the solid lines are the total best fit. $\chi^2$ values 
computed without accounting (``opt'') and accounting (``opt+FIR'') for the 
available 24$\mu$m flux are reported.
If an AGN was detected, a multi-component fit is performed. In these cases, 
the dashed lines are the stellar + starburst dust component, the dotted lines 
represent the torus model and the solid lines are the sum of the two. A unique 
$\chi^2$ value is derived, taking into account all the available data.
}
\label{fig:ib_seds1}
\end{figure}

\end{landscape}

Figure \ref{fig:ib_seds1} shows the spectra and SEDs of IR-peak targets
with spectroscopic redshifts,
the distinction between 4.5$\mu$m- and 5.8$\mu$m-peak being obvious.

Tables \ref{tab:spec_IB_1} and \ref{tab:spec_IB_2} list the basic measured properties of the detected lines for 
IR-peak galaxies, including the equivalent width (EW), full width at half maximum (FWHM, 
corrected for the instrumental resolution), observed
wavelength and identification. In the last columns of Tab. \ref{tab:IB_phot}, spectroscopic 
and photometric redshifts for these targets are listed.

Infrared peakers lie in the fourth quadrant 
of the IRAC color space 
($[3.6-5.8]>0$, $[4.5-8.0]<0$ in the right panel of 
Fig. \ref{fig:ib_r_f24_lacy}; or $[3.6-4.5]>0$, $[5.8-8.0]<0$ in the left panel).
Objects at the boundary between the 
IR-peak and AGN-torus loci (i.e. $[4.5-8.0]\simeq0.0$, $[3.6-5.8]>0$) display mixed properties. 
For example, found here are objects showing 
a power-law like IRAC SED, with a smooth 5.8$\mu$m-peak superimposed (filled circles with crosses).
These can be interpreted as a 5.8$\mu$m peak diluted by 
AGN-torus power-law emission. Other composite galaxies show a clear 
IRAC peak shape, but their restframe UV spectra are dominated by broad-line emission 
(filled circles within open squares). The properties of individual IR-peakers are 
discussed below.

\begin{figure*}[!ht]
\centering
\includegraphics[width=0.9\textwidth]{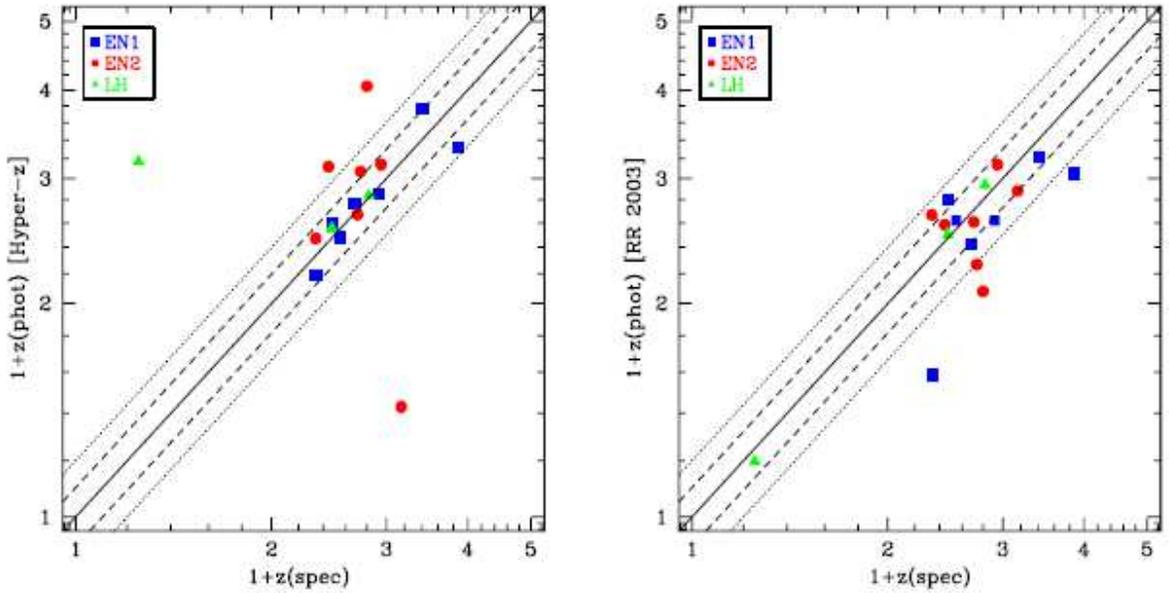}
\caption{Comparison between spectroscopic and photometric redshifts
for IR-peak sources.
{\em Left panel}: results obtained with Hyper-z \citep{bolzonella2000}.
{\em Right panel}: results of the \citet{rowanrobinson2003} code.}
\label{fig:z_spec_phot_IB}
\end{figure*}

Fig. \ref{fig:z_spec_phot_IB} presents the comparison 
between spectroscopic and photometric redshifts for IR-peakers only, 
obtained with the \citet{bolzonella2000} (left hand side) and \citet{rowanrobinson2003}
(right) codes.
The two principal outliers (in the Hyper-z case) are a Lockman Hole source with spectroscopic redshift
of $z=0.249$, which turns out to be a confused object (see below) and a 
broad-line object in ELAIS-N2. In the latter case, the Hyper-z code 
could not find a reliable redshift by adopting a QSO template, because 
of the almost featureless continuum, and then it completely underestimated the 
photometric $z$. The \citet{rowanrobinson2004}  templates
seem to solve these troubles, but underestimate the redshift of a 4.5$\mu$m-peaker 
in ELAIS-N1 (EN1\_202261) with a very red $r^\prime-Z$ color 
(see Fig. \ref{fig:ib_seds1}), which probably was interpreted as 
a deep Balmer break at $z<1$.

\subsection{4.5$\mu$m-peak sources}\label{sect:bump2}

Eight 4.5$\mu$m-peak galaxies were observed. Three of 
these sources have a confirmed spectroscopic redshift between $z=1.30$ and $z=1.45$, with narrow
emission lines (EN1\_202261, EN2\_167372, EN2\_166134). 
The detected features are not enough to distinguish between  
starburst or AGN activity (see Sect. \ref{sect:spec_class}), therefore 
we put these sources in the general ``ELG'' class. For two of these targets
the photometric redshifts agree with the measured spectroscopic 
values (see Tab. \ref{tab:IB_phot}), while for source EN2\_167372 the two 
are not consistent with each other ($z_{phot}=2.120$, $z_{spec}=1.445$).
For this source, IRAC photometry is available only in channels 1 and 2, while both the 5.8 and 8.0
$\mu$m bands have only an upper limit. It is possible that 
the photometric estimate of redshift has been affected by the lack of near-IR 
restframe data.

A fourth 4.5$\mu$m-peak object lies at $z=1.545$ (target EN1\_202260) and is characterized by broad
emission lines, which testify to the presence of a type-1 AGN component contributing to the UV-optical
emission. This object is detected also in the X-rays. The IRAC spectral energy distribution shows 
a red $8.0-5.8$ $\mu$m observed color (see Fig. \ref{fig:ib_seds1}), the
8.0$\mu$m observed flux being likely dominated by hot dust heated by the AGN component.
Photometric and spectroscopic redshift estimates are in agreement, the former having been
obtained with a type-1 AGN template (Polletta et al., in prep.).

\begin{figure*}[!ht]
\centering
\includegraphics[width=0.90\textwidth]{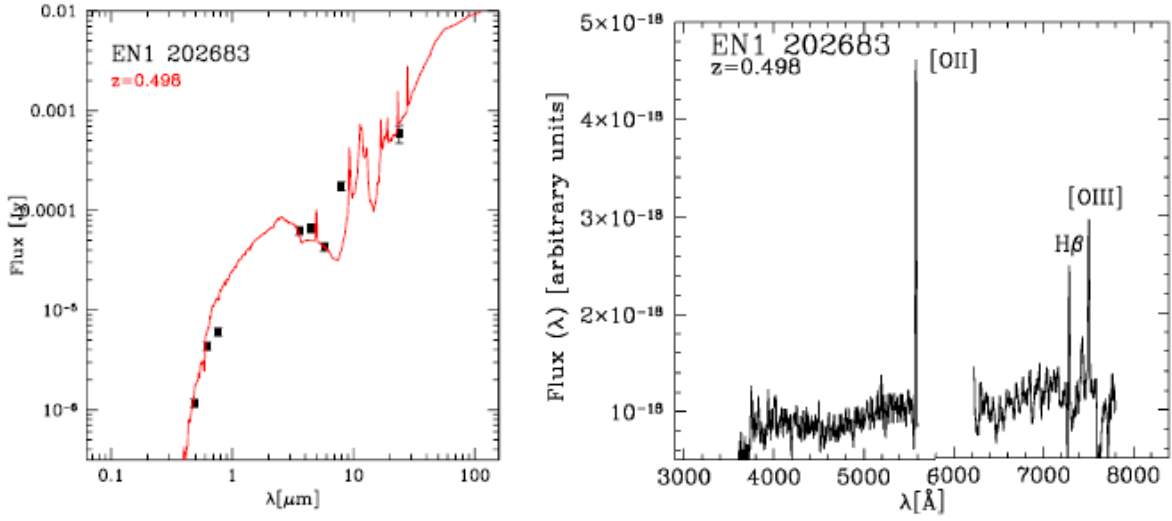}		
\caption{Example of low-redshift starburst galaxy displaying an IRAC
SED which resembles a 4.5$\mu$m peaker.
Note that the 8.0$\mu$m flux density is higher than the 5.8$\mu$m, but
note also that source EN1\_202260 (Fig. \ref{fig:ib_seds1}) has the same 
property but lies at $z=1.545$. The photometric data are compared to the prototypical M82 starburst 
template, normalized to the 5.8$\mu$m band.}
\label{fig:ib2_counterexample}
\end{figure*}

\begin{figure*}[!ht]
\centering
\includegraphics[width=0.45\textwidth]{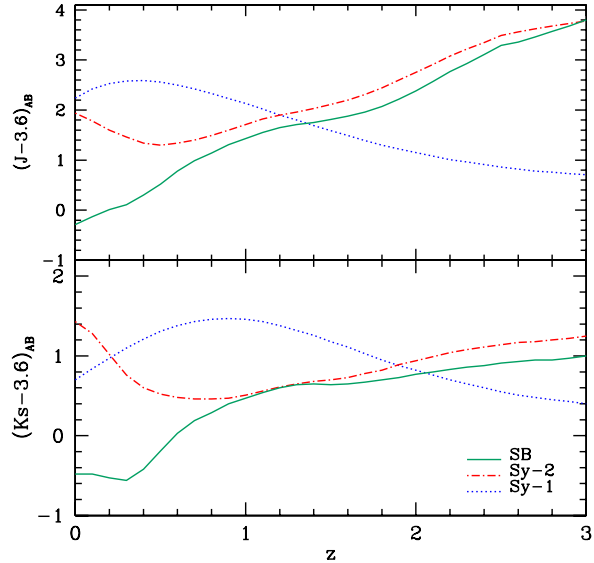}
\caption{Trend of JK-IRAC colors, as a function of redshift, as computed for starburst \citep[M82,][]{silva1998}, 
seyfert-2 \citep[IRAS 19254-7245,][]{berta2003} and seyfert-1 \citep[Mrk231,][]{fritz2006} templates.}
\label{fig:K_col}
\end{figure*}

It is worth noting that a similar IR SED can be observed also in the case of 
starburst galaxies at redshift $z=0.3-0.5$. Actually, in this case a 4.5$\mu$m peak can be produced
by the strong 3.3 $\mu$m Polycyclic Aromatic Hydrocarbon (PAH) feature
lying in IRAC channel-2, while the 8.0$\mu$m flux density is enhanced by mid-IR 
PAH dust features (6.2, 7.7 $\mu$m) at low redshift ($z\le0.5$).
A good example is given by source EN1\_202683 (see Fig. \ref{fig:ib2_counterexample}), 
a starburst galaxy at $z=0.497$. 
An advantage in breaking this kind of aliasing would be 
provided by J, H, K band data. In fact the observed $(J-3.6)_{AB}$ color of a 
typical starburst (e.g. M82) is $\sim0.3$ at $z=0.4$, while it increases to values $>1.5$ 
at $z\ge1.4$. Similarly, the $(K_s-3.6)_{AB}$ values are $\sim-0.4$ and $\sim0.6$
at the same two redshifts (see Fig. \ref{fig:K_col}). 
At $z=0.4$, the 1.6$\mu$m peak lies in the $K_s$ band and the $K_s$ flux is 
brighter than in the IRAC bands. The $K_s$ band is gradually shifted
blueward of the peak, at increasing redshift, and the $(K_s-3.6)_{AB}$ color 
changes sign at $z\simeq0.6$. 

Such low-$z$ 4.5$\mu$m-peak-like sources are not reported in Tabs. \ref{tab:spec_IB_1}
and \ref{tab:spec_IB_2}, since they do not conform to the IR-peak selection (because their
peak is not due to the 1.6$\mu$m feature) and they are easily spectroscopically identified
by using bright optical lines.

On the other hand, another source (EN2\_275226) shows a clean 4.5$\mu$m-peak, 
with $S(8.0)<S(4.5)$ and no rise in IRAC channel 4, but the observed UV 
(restframe) spectrum shows that it hosts a type-1 AGN as well. This object lies at 
$z=1.710$, in perfect accordance with the 4.5$\mu$m-peak selection
and the photometric redshift estimate. It is likely that the AGN component
dominates the UV restframe spectrum, while the near-IR SED is powered mainly 
by star light.

Finally, source LH\_572243 lies at $z=1.820$, having narrow
lines detected in its spectrum. 
A bright, narrow {\sc Civ} ($\lambda=1549$ \AA) emission
line, without any P-Cyg profile, is detected, testifying to the presence of a type-2 AGN nucleus
\citep[e.g.][]{farrah2005}, which is also confirmed by X-ray data.

The remaining 4.5$\mu$m-peak targets (LH\_245782 and LH\_576281) don't have any spectroscopic redshift confirmation, 
although their continuum emission is detected by LRIS.

As a whole, 6 out of the 8 4.5$\mu$m-peakers have a robust
spectroscopic redshift, as derived from emission lines. Considering that LH\_245782 was included in one of the two masks 
observed at high airmass, this translates into a 87.5\% successfull detection for 4.5$\mu$m
peakers. Besides the presence of AGN components, our photometric estimates of redshift were 
consistent with the actual spectroscopic evidence in 80\% of the cases, while 
for the remaining poor photometry is the main cause of discrepancy.

\subsection{5.8$\mu$m-peak sources}\label{sect:bump3}

Our masks include 27 5.8$\mu$m-peak targets, of which only 10 have a confirmed spectroscopic
redshift. Thirteen out of the 17 without spectral lines detected lie in the Lockman Hole field,
whose observing conditions were not optimal for optically-faint objects.
The remaining four are all faint targets, having $r^\prime\ge23.5$.

In the 5.8$\mu$m-peaker sample, we distinguish between sources having a clear
stellar 1.6$\mu$m peak detected in IRAC bands (e.g. object EN2\_172324, see Fig.
\ref{fig:ib_seds1}) and sources with a steep infrared SED (e.g. target
EN1\_279954, Fig. \ref{fig:ib_seds1}). The latter are likely to be composite sources, whose near-IR
emission is not only due to low-mass stars, but also to an AGN component. 
The 1.6$\mu$m restframe peak is produced by the stellar component, but it is
diluted by the AGN emission, which reddens the $5.8-8.0$ $\mu$m color.
These sources are characterized by a very red $5.8-4.5-3.6$ $\mu$m slope, 
with flux ratios of $S(5.8)/S(4.5)\simeq1.4-1.8$ and $S(4.5)/S(3.6)\simeq1.3-1.5$.

This group consists of targets EN1\_205467, EN1\_279954 and EN2\_165986. 
For the first one we were not able to derive any spectroscopic redshift, while the 
other two have very bright, broad ultraviolet emission lines (see Tabs. \ref{tab:spec_IB_1} and \ref{tab:spec_IB_2}),  
confirming the presence of a type-1 AGN contributing to the optical observed fluxes.
The two sources turn out to be at redshifts $z=2.409$ and $z=2.163$ respectively, 
consistent with the 1.6$\mu$m SED feature lying in the 5.8$\mu$m IRAC band. The photometric
redshift estimate for this kind of source is carried out by adopting an AGN-like
template (Polletta et al., in prep.), which leads to a 50\% consistency with
the actual spectroscopic measure (see Tab. \ref{tab:IB_phot}). 
This problem can be ascribed mainly to the 
lack of sharp features in the broad-band SEDs of these sources.

Sources EN1\_282078 and EN2\_273717 show a bluer slope, with 
$S(5.8)/S(4.5)\simeq1.2$ and $S(4.5)/S(3.6)\simeq1.1-1.3$ and a smooth
5.8$\mu$m peak. These sources turn out to have broad
emission lines as well, at lower redshift, namely $z=1.685$ and $z=1.800$ respectively.
The AGN contribution to the NIR SED of these sources is likely lower than in the previous case, 
but still non-negligible and clearly visible in the UV-optical domain (restframe).
The inconsistency of the spectroscopic redshift value and the 5.8$\mu$m-peak 
selection --- which implies that the restframe 1.6$\mu$m stellar feature is shifted 
to the IRAC channel 3 at $z>1.8$ --- is discussed below, and interpreted 
with a non-negligible AGN contribution to IRAC fluxes (Sect. \ref{sect:IB_SED}).

Also in the case of source LH\_574364, the IRAC data define a smooth 5.8$\mu$m peak, but the
$S(8.0)/S(4.5)$ flux ratio is smaller than unity in this case (while it was larger than 1 in the two
classes described above). Narrow Mg{\sc ii} and Fe emission lines are detected. The latter
suggests the presence of a type-2 AGN. 
The smoothness of the peak suggests that this galaxy might lie at
$z<1.5$ with the 1.6$\mu$m feature falling shortward of the 5.8$\mu$m band center. In fact the
spectroscopic redshift of this galaxy turns out to be $z=1.474$, 
consistent with the photometric estimate.

The target EN1\_339960 is formally a 5.8$\mu$m-peaker, but its channel 2 and 3 fluxes are 
comparable, defining a very broad peak in the IRAC SED. The observed spectrum is 
dominated by a very strong {\sc Civ} emission with a $\sim2000$ $[$km s$^{-1}]$ width, at $z=1.475$,
putting it in the AGN-1 spectroscopic class.

Three galaxies have a confirmed redshift between 1.7 and 2.0, narrow Ly$\alpha$ and other emission
lines and SED typical of high-redshift starbursts, i.e. EN1\_342445, EN2\_11091 and
EN2\_172324. The latter shows a bright {\sc Civ} narrow line, with no P-Cyg profile, 
that we interpret as produced by a type-2 AGN. A reddened torus contributes also to the 
IRAC SED, shifting the peak to the 5.8$\mu$m band (see Section \ref{sect:IB_SED}).
For the other two sources, not enough spectral features are detected, therefore they are 
simply classified as ``emission line'' galaxies.

The highest redshift IR-peaker detected is a 
a $z=2.866$ source, EN1\_340451, lying at the upper redshift limit of the
5.8$\mu$m-peak selection.
This is a very faint optical source, with bright IRAC emission 
and a weak 24$\mu$m flux. We detect Ly-$\alpha$, He{\sc ii} ($\lambda=1640$ \AA)
and {\sc Ciii]} ($\lambda=1909$ \AA) narrow lines. The wavelength coverage of 
the data has a gap in the range $\lambda=5500-6150$ \AA, where {\sc Civ} ($\lambda=1549$ \AA) 
would fall, but the presence of He{\sc ii} (with a ionization energy of 54.4 eV, four times 
that of {\sc Hi}) suggests the presence of a type-2 AGN component.

\begin{figure*}[!ht]
\centering
\includegraphics[width=0.93\textwidth]{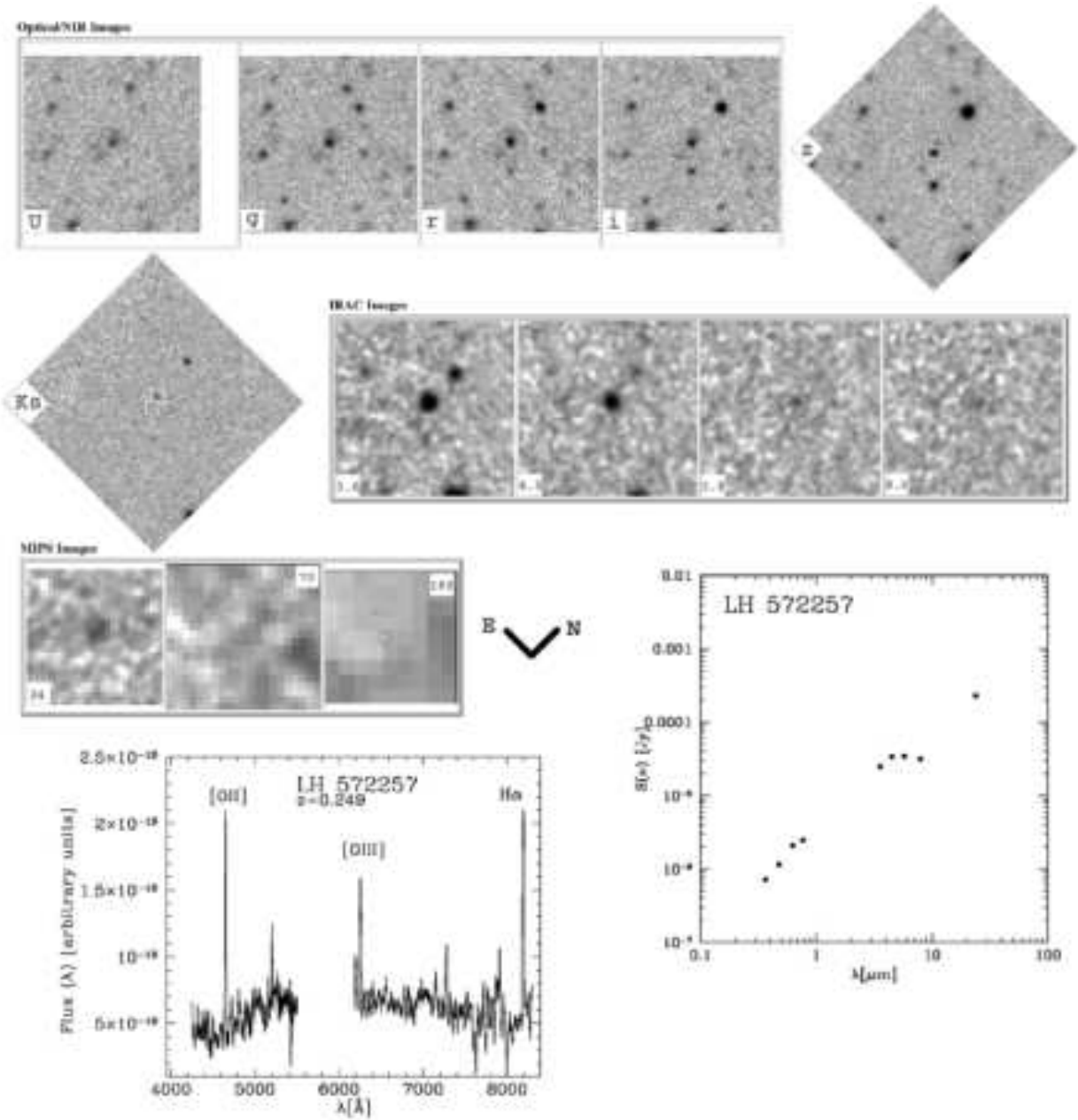} 
\caption{The 5.8$\mu$m-peak target LH\_572257 at $z=0.249$. 
White squares are drawn on the $z$ and $K_s$ band stamps, 
in order to guide the reader in finding the two physical components 
that contribute to the observed SED (see text for more details). 
Stamp size is 0.01 deg.}
\label{fig:572257}
\end{figure*}

Finally, it is important to highlight the presence of one outlier: the low redshift interloper 
LH\_572257, at $z=0.249$.
This object is an X-ray source, with soft ($0.3-2.5$ keV), hard ($2.5-8.0$ keV) 
and broad ($0.3-8.0$ keV) fluxes of 11.01, 26.71 and $35.15\times10^{-16}$ $[$erg
cm$^{-2}$ s$^{-1}]$.
By examining the optical and IRAC postage stamp images in Fig. \ref{fig:572257}, one can notice that
optically the source seems to consist of two different components: a point-like object (likely the
X-ray low-$z$ object) and a fuzzy galaxy, $\sim 1.5{\arcsec}$ to the north. 
It is possible that the latter dominates the IRAC SED and
effectively is a high-redshift 5.8$\mu$m-peak galaxy lying underneath a low-$z$ source, which instead
dominates the optical fluxes. This is confirmed by the Ks band image, where the
optically-faint galaxy dominates and the optically-brighter object is significantly fainter.
Couples of white squares are 
over plotted on the $z$-band and $K_s$-band images, centered on the 
two distinct components.

\subsection{Sources with no emission line detection}\label{sect:ib_nolines}

Apart from objects in the Lockman Hole, for which the observing conditions
were not optimal (high airmass), there are four 
IR-peakers with no line detection in the ELAIS fields. 
In all cases continuum emission is detected; in Tabs.  \ref{tab:spec_IB_1} and \ref{tab:spec_IB_2}
we report the wavelength range covered by the detected continuum for each source.

\begin{figure*}[!ht]
\centering
\includegraphics[width=0.9\textwidth]{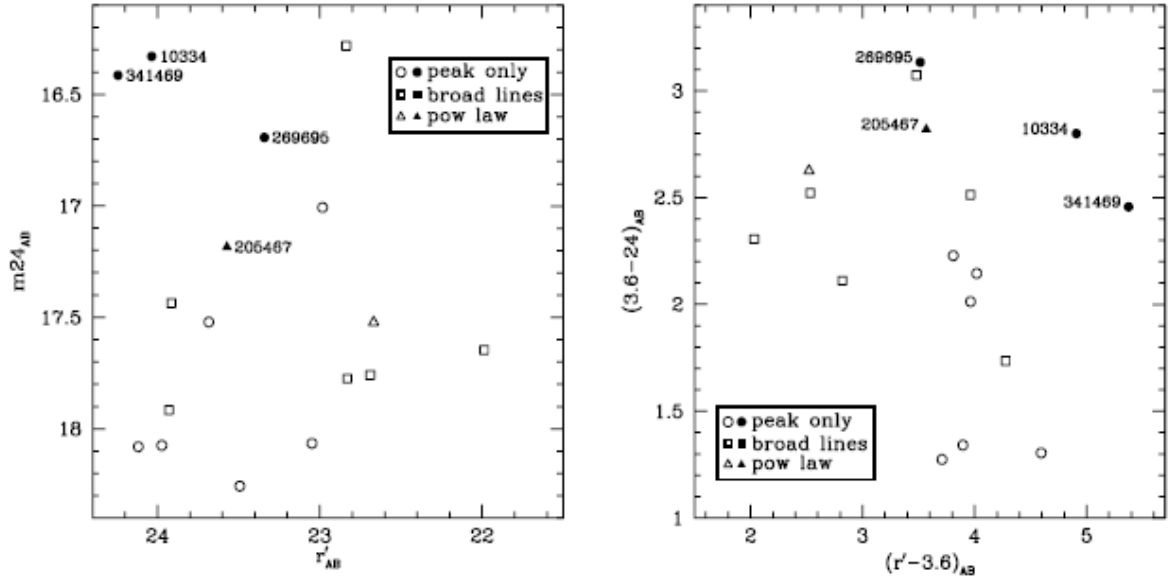}
\caption{Distribution of IR-peak targets in the optical-IR color space
(ELAIS-N1 and N2 only). Open symbols belong
to sources with a spectroscopic redshift, filled ones represent objects with no
spectral features detected.}
\label{fig:IB_not_det}
\end{figure*}

The IR-peakers without line detection lie at the faint end of the $r^\prime$ magnitude distribution
of our targets ($r^\prime>23.5$, see Tab. \ref{tab:IB_phot}). Nevertheless, for other comparably faint IR-peak sources 
a measure of the spectroscopic redshift has been possible.
In Figure \ref{fig:IB_not_det} (left panel) the $r^\prime$ and 24$\mu$m AB magnitudes 
of the ELAIS-N1, N2 IR-peakers are compared. The sources without line detection turn out to be 
those with brighter 24$\mu$m fluxes, of the order of $0.5-1.0$ mJy.

The plot on the right shows that these galaxies are also 
those with the brighter mid-IR excess, i.e. the reddest $(3.6-24)$ $\mu$m color.
The two reddest ones lie close to a broad-line IR-peaker.
One (EN1\_205467) of the two shows a power-law IRAC SED (plus 5.8$\mu$m-peak)
and is a point-like object on optical images; its photometric redshift is $z=0.320$, but 
it is not reliable because of the almost featureless broad-band SED.

The second one (EN2\_269695) is a fuzzy galaxy on optical images
and shows a sharp 5.8$\mu$m peak. The photometric redshift is 
$z=1.800$, which would allow UV emission lines (Ly$\alpha$, Mg{\sc ii})
to be detected in the covered spectral range (see Tab.\ref{tab:spec_IB_1}).
The huge mid-IR luminosity suggests that this is a powerful starburst, 
heavily extinguished by dust, which would also explain the 
lack of detected emission lines.

The remaining two objects (EN1\_341469 and EN2\_10334) are the two optically-reddest 
in the IR-peaker category, with $(\textrm{r}^\prime-3.6\mu\textrm{m})>4.9$
(in AB units, see Fig. \ref{fig:IB_not_det}).  
These two sources are also the two brightest 24$\mu$m emitters, with 
$S(24)>1.0$ mJy.

The photometric redshifts are $z=1.54$ and 2.01 respectively;
by adopting a typical starburst template (e.g. M82), we derive 
bolometric infrared ($8-1000\ \mu$m) luminosities of $5\times10^{12}$ L$_\odot$ 
for both sources, which would imply an SFR in excess of 800 M$_\odot$/yr
\citep{kennicutt1998}, if powered only by star formation.
Such level of activity would produce a Ly$\alpha$ emission of 
$2.5\times10^{11}$ L$_\odot$.

Unfortunately, for EN1\_341469 no blue-arm spectrum is available
and we cannot test a possible detection of Ly$\alpha$.
As far as EN2\_10334 is concerned, at a redshift $z=2.01$, that Ly$\alpha$ luminosity corresponds 
to a flux of $S(\textrm{Ly}\alpha)=3.1\times10^{-14}$ $[$erg cm$^{-2}$ s$^{-1}]$
and the line would fall at $\lambda=3660$ \AA.
The continuum of source EN2\_10334 is detected 
at a 2$\sigma$ level with a flux of $2.0\times10^{-18}$ $[$erg cm$^{-2}$ s$^{-1}$ \AA$^{-1}]$.
Adopting an intrinsic FWHM of 10 \AA,
in order not to detect the line at a 5$\sigma$ level, 
the source must be extinguished by $\textrm{A}(\textrm{Ly}\alpha)=5.75$
magnitudes, corresponding to $\textrm{A}_V\simeq2$ 
(using the Calzetti et al. \citeyear{calzetti1994} extinction law).

In the case that a type-2 AGN plays a non negligible role in the 
emission mechanism, then these numbers would 
significantly change.

\subsection{Modeling of observed SEDs}\label{sect:IB_SED}

The spectral energy distributions (SEDs) of the IR-peak galaxies with spectroscopic redshifts are
analysed here, by means of multi-component fitting.

For sources that do not show any AGN signature in their spectra or broad-band
SEDs, we adopt the technique described in \citet{berta2004} of 
mixed stellar population synthesis.
The observed SEDs are reproduced by combining simple stellar populations (SSPs)
of solar metallicity, built on the basis of the Padova 1994 isochrones \citep[for more details on the SSP library 
see][]{berta2004}. Each phase in the SSP history 
is extinguished by a different amount of dust, according to age-selective
extinction \citep{poggianti2001}. 
Since disc populations are on average affected by a moderate $A_V$ 
\citep[$<1$ mag, e.g.][]{kennicutt1992}, the maximum allowed absorption for 
stars older than 1 Gyr is $A_V=0.3-1.0$ magnitudes. For younger populations
the color excess gradually increases, but is limited to $A_V\le5$.

The 24$\mu$m flux is included in the fit and it is
used for constraining the amount of dust extinguishing young stars in the ongoing starburst.
The energy absorbed by dust at UV-optical wavelengths is reprocessed to the mid- and far-IR domain, 
by means of a prototypical starburst template. 

An M82 template was adopted, 
built combining the \citet{silva1998} model and 
the \citet{foerster2001} observed mid-IR spectrum.
Local ultra-luminous sources (L$_{IR}>10^{12}$ L$_\odot$)
can be characterized by deep silicate 10$\mu$m self-absorption and 
``cold'' SEDs (e.g. Arp220). Similar templates 
equally show the usual stellar 1.6$\mu$m peak.

Nevertheless, increasing observational evidence exists that high-$z$
IR-peakers detected in the mid-IR resemble the M82 prototype.
Spitzer mid-IR IRS spectra of $z\simeq1.9$ IR-peak galaxies \citep{weedman2006} 
are dominated by bright PAH features and lack silicate 10$\mu$m self-absorption.
\citet{rowanrobinson2005} studied and classified the SEDs of SWIRE sources over 6.5 deg$^2$
in the ELAIS-N1 field. These authors find that M82-like starbursts are 3 times 
more numerous than colder Arp220-like objects. Based on the observed 
redshift distribution and on number counts modeling, they also infer that 
this ratio is even higher at $z\sim1.5-2.0$.
Lonsdale et al. (in press) performed 1.2 mm (250 GHz) observations of SWIRE 24$\mu$m-selected 
(H)ULIRGs with the Max Plank Millimeter Bolometer \citep[MAMBO,][]{kreysa1998} on the IRAM/30m telescope.
As a result, they find that the 1.2mm/24$\mu$m flux ratio of these sources resembles that 
of M82, lower than for an Arp220-like population, and lower than what found for sub-mm 
selected galaxies (SMGs). It is worth noting, in fact, that their selection 
(similar to ours) favors 24$\mu$m-bright systems, instead of mm-bright objects.
All of these recent findings drove the choice of the M82 template, as stated above.

As a further check,
we have attempted stacking of far-IR (70 and 160 $\mu$m) SWIRE images at the position of 
the observed IR-peakers, but unfortunately no signal was detected on the 
stacked frames.

Also the detected spectroscopic features have been included in the fitting procedure, 
in order to provide additional constraints on the star formation history of 
these sources. 

Table \ref{tab:fit_sb} reports the results of this analysis: for each source, the 
best fit stellar mass and ongoing star formation rate (SFR) are listed. The SFR is computed
as an average on the last $10^8$ yr in the life of the galaxy, in order to be comparable to the \citet{kennicutt1998}
calibrations. We also report the derived extinction, as  averaged over the whole galaxy life and over the last 
$10^8$ yr, as well as the bolometric IR ($8-1000$ $\mu$m) luminosity, as computed 
with the adopted template. Finally, 3$\sigma$ ranges, as computed from the exploration of the 
parameters space, are listed within parentheses. A detailed description of degeneracies is provided in \citet{berta2004}.
The average number of SSPs effectively involved in the fit is 3-4, 
depending on redshift.

In Fig. \ref{fig:ib_seds1} we show the fit to the observed SEDs.
The blue-dashed and red-dotted lines represent the contributions of young (age $<10^9$ yr) 
and old (age $\ge10^9$ yr) stars to the modeled SEDs. The green solid line is the total
emission in the optical, while the long-dashed cyan line longward of 5$\mu$m (restframe)
is the M82 template. The integral of the template between 8 and 1000 $\mu$m reproduces the 
absorbed energy in the UV-optical. On the plots, we report the $\chi^2$ values (not reduced),
the stellar mass and the fraction of mass in young/old stars.

The IR-peak sources fitted in this way turn out to be powered by strong starburst activity,
with SFRs reaching 400 M$_\odot$/yr and IR luminosities exceeding $10^{12}$ L$_\odot$ in the most
powerful cases.
The observed SEDs are overall well reproduced over the whole spectral domain 
from the U band to 24 $\mu$m, with reduced $\chi^2$ between 1 and 2 
\citep[see][ for a discussion on high $\chi^2_\nu$ values]{berta2004}.

The LRIS spectrum of source LH\_572243 shows a weak, but clear, {\sc Civ} emission, with no 
P-Cyg profile, which we interpreted as due to AGN activity. The stellar synthesis
fit, however, reproduces the observed SED without the need of any type-2 AGN component. 
Similarly, we classified source EN1\_340451 as a type-2 AGN, on the basis 
of the He{\sc ii} narrow line detected with Keck, but its broad-band SED is 
well fitted by a stellar model, with no need for additional components. Thus
this source is best fitted by a moderate starburst (SFR$=56.1$ M$_\odot$/yr) hosted in 
an extremely massive galaxy with M$=6.69\times10^{11}$ M$_\odot$. 

A detailed analysis of the stellar mass function of IR-peak sources is being 
carried out for the SWIRE survey, and is deferred to Berta et al. (in prep.),
which will take into full account the spectroscopic results presented here.

As far as sources with an AGN spectral classification are concerned, a different 
fitting procedure was adopted. We reproduced the observed datapoints by means 
of the combination of a simple stellar population and a torus template \citep{fritz2006}.
The purpose, in this case, is to show how the detected IR-peak can be reproduced with 
multiple components, when the spectroscopic redshift is not fully consistent with what was expected for 
a pure stellar 1.6 $\mu$m peak.

We have therefore combined the \citet{fritz2006} torus library, with the 
simple stellar population library used before \citep{poggianti2001,berta2004}.
The best fit is sought by $\chi^2$ minimization. The stellar component consists of 
one SSP, extinguished by a varying amount of dust. Again, extinction is also constrained through
a far-IR starburst template. In combination to stars, the torus AGN emission is added, in 
order to fit the observed data.

The torus library by Fritz et al. spans several geometries of the toroidal dust distribution
around the central AGN nucleus, varying the ratio between outer and inner radii ($R_{out}/R_{in}=20-300$),
and the aperture angle of the torus (measured starting from the equatorial plane, $\Theta=40^\circ-140^\circ$).
We limit our analysis to a uniform dust distribution in the torus, because
we do not have sufficient data points to constrain the entire library and because our purpose here is
to get a general requirement on the level of AGN contribution to the mid-IR SED.
The optical depth at the equator covers the range $\tau=0.1-10.0$ at 9.7 $\mu$m. 
The spectrum emitted by the central engine is modeled with a 
broken power-law $\lambda\, L(\lambda)\propto\lambda^\alpha$, with indexes $\alpha=1.2,\ 0,\ -0.5$
in the ranges $\lambda=0.001-0.03, \ 0.03-0.125, \  0.125-20$ $\mu$m respectively.
Two different sets of models are included, differing for the UV-optical-IR
slope of this power law, being $\alpha=-0.5$ or $\alpha=-1$.
See \citet{fritz2006} for more details of this model.

The results of this AGN+stars fit are in Tab. \ref{tab:fit_agn} and plotted 
in Fig. \ref{fig:ib_seds1}. The green dashed line
represents the stellar component (including starburst dust), the blue dotted line 
is the AGN contribution to the global SED (red solid line).

The details of the best fit models are subject to strong degeneracies and are limited
by the use of one single SSP, instead of a sophisticated model like in the starburst case.
A unique best fit can not be achieved due to 
the limited photometry, therefore 3$\sigma$ ranges for the parameters are reported in Tab. \ref{tab:fit_agn}, 
instead of best fit values. These fits have the illustrative purpose 
to point out how the shape of the broad band SED of IR-peak galaxies
with AGN detection in the UV restframe spectra can be simply explained by two different physical
components.

According to the unified scheme for AGN emission \citep{antonucci1985}, in the UV type-1 features emerge when the viewing angle 
does not intercept the dusty torus. At longer wavelengths, stellar and AGN spectral energy distributions are complementary,
with the stellar component peaking around 1 $\mu$m and dominating the optical spectrum, while the torus 
warm dust emission increases longward and dominates the 3-10 $\mu$m range (see, for example, source EN1\_202260).
The warm component of the torus, coming from the inner regions closer to 
the central engine, can significantly contribute to the IR-peak itself, modifying its shape
and apparently shifting it to a wavelength longer than 1.6$\mu$m (restframe).
A good example of this effect is source EN1\_282078, for which the IR-peak 
is detected in the restframe K band.

When a type-2 AGN is present, which in fact happens only in three of our sources, 
the UV-optical SED can be easily fitted by stars alone. In two cases (EN1\_340451 and 
LH\_572243) the whole SED, up to the mid-IR is reproduced with a simple starburst
model (see above). Only in one case (EN2\_172324) is
a warm dust component needed in order to explain the near-IR fluxes.
Nevertheless, in this latter case, our code cannot fully reproduce the 
observed shape of the IR-peak, even including the type-2 AGN.

Other effects might explain the observed SED of EN2\_172324.
Dust in AGB stars can significantly modify the shape of the IR-peak, changing
its colors. The feature flattens, becoming bluer on the blue side and redder on the red side 
\citep[see][]{piovan2003}, and the H$^-$ feature is smoothed.
Nevertheless these effects cannot explain a shift in the redshift of the peak of $\Delta z=0.4$,
as apparently observed in source EN2\_172324.

At the observed redshift ($z=1.739$), both Br$\gamma$ and Pa$\alpha$ fall in the 5.8$\mu$m 
IRAC channel and could give a significant contribution to the observed flux, if very strong.
Obviously, as a final source of uncertainty, photometry plays an important role.

Finally, blind tests on the sources listed in Tab. \ref{tab:fit_sb} (i.e. with no evidence 
for an AGN in their spectra) were carried out, using the multi-component approach.
The fit confirm that a possible AGN would contribute less than 2\% to the IRAC fluxes of these objects.

\section{Other interesting sources}

The multi-object mask geometry, and the low density of IR-peakers on the sky, allowed 
us to target several SWIRE/Spitzer sources having 
peculiar properties or multiwavelength counterparts over a wide 
fraction of the electromagnetic spectrum.

The properties of some of these interesting sources are briefly 
described here, while a detailed analysis is deferred to 
subsequent papers.

\subsection{High-redshift AGNs}

The observed masks include three type-1 AGNs at redshift $z>2.5$.
Two of these AGNs (EN1\_282051 and EN2\_274735) were selected as $z\sim3$ QSO candidates  
by their flux decrement in the U-band.  We further required a red  
IRAC color ($m[3.6]-m[4.5] > -0.15$) which eliminates contamination from  
main sequence stars (Siana et al. 2007, in prep.).

The third object (EN1\_202756) did not have any particular priority;
it would have been selected as a $z\sim3$ QSO but since it is
fainter ($g^\prime>23.42$, Vega), the U band depth was insufficient to provide a red  
enough $U-g^\prime$ limit for selection.  This AGN is not detected at 24$\mu$m, 
nor in IRAC channels 3 and 4, in the SWIRE survey.

\begin{figure*}[!ht]
\centering
\includegraphics[width=0.80\textwidth]{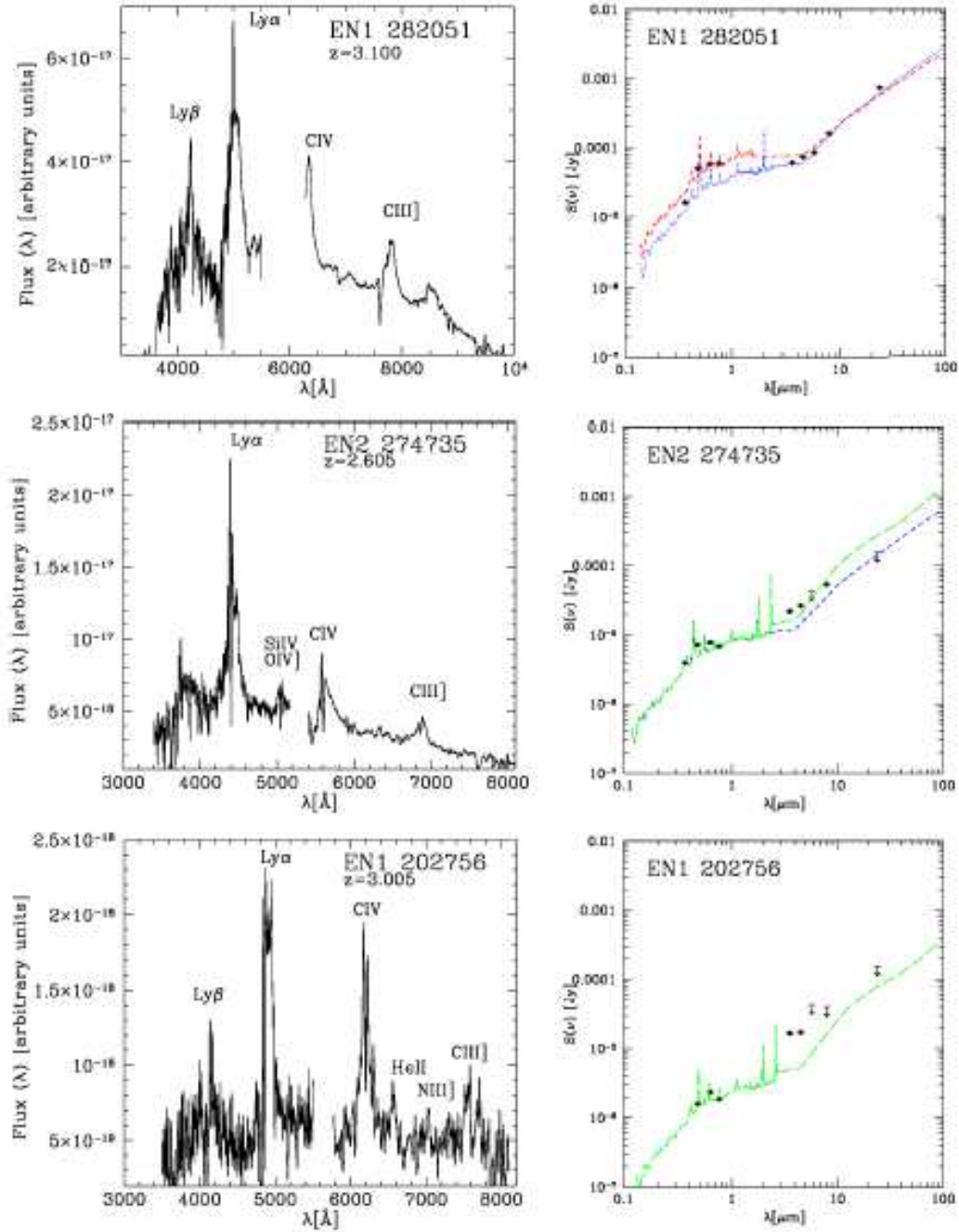}
\caption{Spectra and SEDs of high-redshift QSOs. The observed 
photometry is compared to three different templates. The standard template (dotted lines)
is built with the optical composite quasar spectrum from the Large Bright 
Quasar Survey \citep{brotherton2001} and the average SED of SWIRE quasars \citep{hatziminaoglou2005}.
The reddest (long-dashed lines) and bluest (short-dashed) 
differ from it only in in their IR/optical luminosity ratio \citep{polletta2006}.}
\label{fig:QSOs}
\end{figure*}

EN1\_282051 is a bright quasar, with a 24$\mu$m flux of 740 $\mu$Jy, lying at 
redshift $z=3.1$. Four broad emission lines were detected: Ly$\beta$, Ly$\alpha$,
{\sc Civ} ($\lambda=1549$ \AA) and {\sc Ciii]} ($\lambda=1909$ \AA). 
Figure \ref{fig:QSOs} shows the observed spectrum and the SED of this target.
The latter has been superimposed with two different QSO templates differing 
in the FIR/optical luminosity ratio. The optical part of the templates belongs to 
the composite quasar spectrum from the Large Bright Quasar Survey \citep{brotherton2001}
while the infrared section was obtained as the average SED of SWIRE quasars 
\citep{polletta2006,hatziminaoglou2005}.
A detailed analysis of this object is being carried out by Siana et al. (in prep.).

Five broad emission lines were detected for target EN2\_274735, 
Ly$\alpha$, Si{\sc iv,Oiv]} ($\lambda=1400$ \AA), {\sc Civ} 
($\lambda=1549$ \AA) and {\sc Ciii]} ($\lambda=1909$ \AA),
at redshift $z=2.605$, with several intervening systems producing 
absorption lines.
This object is a optically-bright quasar that 
is not detected in the SWIRE 24$\mu$m survey (at the 200 $\mu$Jy limit) and 
has a FIR/optical color redder than in the previous case (see Fig. \ref{fig:QSOs}). 
Further analysis is deferred to Siana et al. (in prep.). 

Finally, EN1\_202756 is a faint quasi stellar object, with a $r^\prime$ magnitude 
of 22.95 (AB), detected only in the 3.6 and 4.5 $\mu$m channels, with a 16.65 and 17.12
$\mu$Jy flux respectively. Six emission lines are detected at $z=3.005$:
Ly$\beta$, Ly$\alpha$, {\sc Civ}, He{\sc ii} ($\lambda=1640$ \AA), {\sc Niii]} 
($\lambda=1750$ \AA) and {\sc Ciii]}. The observed colors are redder than 
the QSO templates adopted (see Fig. \ref{fig:QSOs}).

\subsection{X-ray sources}

Two masks were centered on the Lockman Hole and ELAIS-N1 regions
which had been observed in the X-rays by Chandra 
\citep{polletta2006,manners2003,manners2004,franceschini2005}.

In the Lockman Hole, three X-ray sources were put on a slit, a 4.5$\mu$m-peaker 
(already discussed in Sect. \ref{sect:bump2}), the troublesome 5.8$\mu$m-peak 
object LH\_572257, which turned out to be a low-redshift confused interloper 
(see Sect. \ref{sect:bump3}), and finally the low-redshift ($z=0.355$)
type-1 AGN LH\_575325, discussed in Wilkes et al. (in prep.).

As far as ELAIS-N1 is concerned, four X-ray sources have a confirmed 
spectroscopic redshift. Two sources have typical type-1 AGN spectra
(EN1\_204120 and EN1\_202260, $z=1.475,\ 1.545$), 
consistent with the classification by
\citet{franceschini2005}; EN1\_203962 is a type-2 AGN at $z=0.874$, 
while finally EN1\_201165 has a starburst spectrum with {\sc [Oii]} and 
Ca{\sc ii}-HK detected at $z=0.762$. \citet{franceschini2005} classify both
targets as Seyfert-2 galaxies, although the latter (target 92 in their work)
shows a lack of X-ray photons with respect to the AGN prediction.

\section{Discussion}

When neither broad lines, nor type-2 lines are present in the spectra of the observed 
IR-peakers, 
a pure stellar spectro-photometric synthesis was performed \citep{berta2004}.
Seven sources satisfy these requirements; their luminosities are 
in (or close to) the ULIRG regime ($L_{IR}\ge10^{12}$ L$_\odot$), and their bright IR fluxes 
turn out to be powered by strong obscured starburst activity. The median rate of star formation 
is $\sim90$ $[$M$_\odot$/yr$]$; the median extinction of the stars in the starburst (age $\le10^8$)
is $A_V=$1.65 magnitudes (see Tab. \ref{tab:fit_sb}). 

The host galaxies of these starbursts are extremely massive, 
$M(\star)=1-6\times 10^{11}$ M$_\odot$, at redshifts $z=1.3-2.8$.
Consequently, the median derived timescale for star formation 
$t_{SF}=M(\star)/SFR$ turn out to be $2.6\times10^9$ years,
requiring many such episodes of star formation in order to 
form the whole assembled mass (provided that the typical 
duration of a starburst episode is $\sim 10^8$ yr).
The two most active objects, with $SFR$ of the order of 500 M$_\odot$/yr, 
have $t_{SF}\sim2-3\times10^8$ yr, fast enough to form the bulk of the
total stellar mass in one single extreme burst of star formation.

A different approach to SED fitting is followed, when dealing with sources that show 
AGN signatures in their spectra or SEDs (seven out of 16 targets). In this case, 
we combine a single SSP and a torus model \citep{fritz2006}.

The results show how warm dust from an AGN torus can 
be a significant contributor to IRAC fluxes, especially to channels 3 and 4, 
for example providing a fraction $>50$\% of the total emission at 8.0$\mu$m.
It is very interesting to point out that in these cases the AGN component
not only dilutes the infrared stellar peak, but also produces an apparent
shift of the peak to longer wavelengths. This effect is basically due to the 
shape of the Planck emission of warm dust at temperatures of few to several hundreds 
Kelvin. The presence of an AGN is sufficient to explain the 
inconsistency between spectroscopic and photometric redshifts of
IR-peakers. 

In type-1 objects, the AGN component dominates the UV restframe emission, 
producing bright broad lines in the observed spectra; in the optical-near-IR 
domain stars and torus SED are complementary, and stars provide the bulk of 
the observed fluxes. The torus component again emerges at longer wavelengths, 
in the near-mid infrared, with warm dust from the inner regions 
dominating between 3-10 $\mu$m restframe.

The best fit solutions preferentially require a small torus distribution of dust 
around the central AGN, with $R_{out}/R_{in}=30-100$. 
The AGN component dominates in the IRAC observed frame, but the fraction of flux due to the AGN
decreases at longer wavelengths. In more than 50\% of cases, the torus usually does not significantly contribute 
to MIPS fluxes (e.g. at 24$\mu$m) and the contribution to the 
total IR ($8-1000$ $\mu$m) luminosity is $\sim15\%$ for the majority of sources (see Tab. \ref{tab:fit_agn}). 
Therefore the mid-IR spectrum of IR-peak galaxies 
is expected to be characterized by bright PAH features.
\citet{weedman2006} confirm the presence of bright 7-13 $\mu$m PAHs 
in the IRS spectra of optically-faint 5.8$\mu$m-peak galaxies, with no evidence of dilution by AGN torus dust in 
90\% of the examined cases.
This suggests that the mid-IR emission of Weedman's galaxies is dominated by starburst activity.
Nevertheless, their sources lie at $z\simeq1.9$, on the lower bound of the 5.8$\mu$m selection;
even in the case of these optically-faint sources, it is possible that 
an AGN (torus) component provides a non negligible contribution to IRAC fluxes, 
without being identified in the mid-IR.

All the sources with type-1 AGN broad lines detected in the Keck spectra require a
non negligible contribution of torus warm dust to their IRAC SEDs (see Tab. \ref{tab:fit_agn}),
while two objects with type-2 AGN classification can be fitted with a stellar component only,
with no needs of any torus to reproduce the observed IRAC fluxes.
Conversely, all sources showing a significant excess in the IRAC domain, with respect to 
pure stellar emission, show AGN signatures in their UV-optical restframe spectra (either type-1 or 2).

\begin{figure*}[!ht]
\centering
\includegraphics[width=0.9\textwidth]{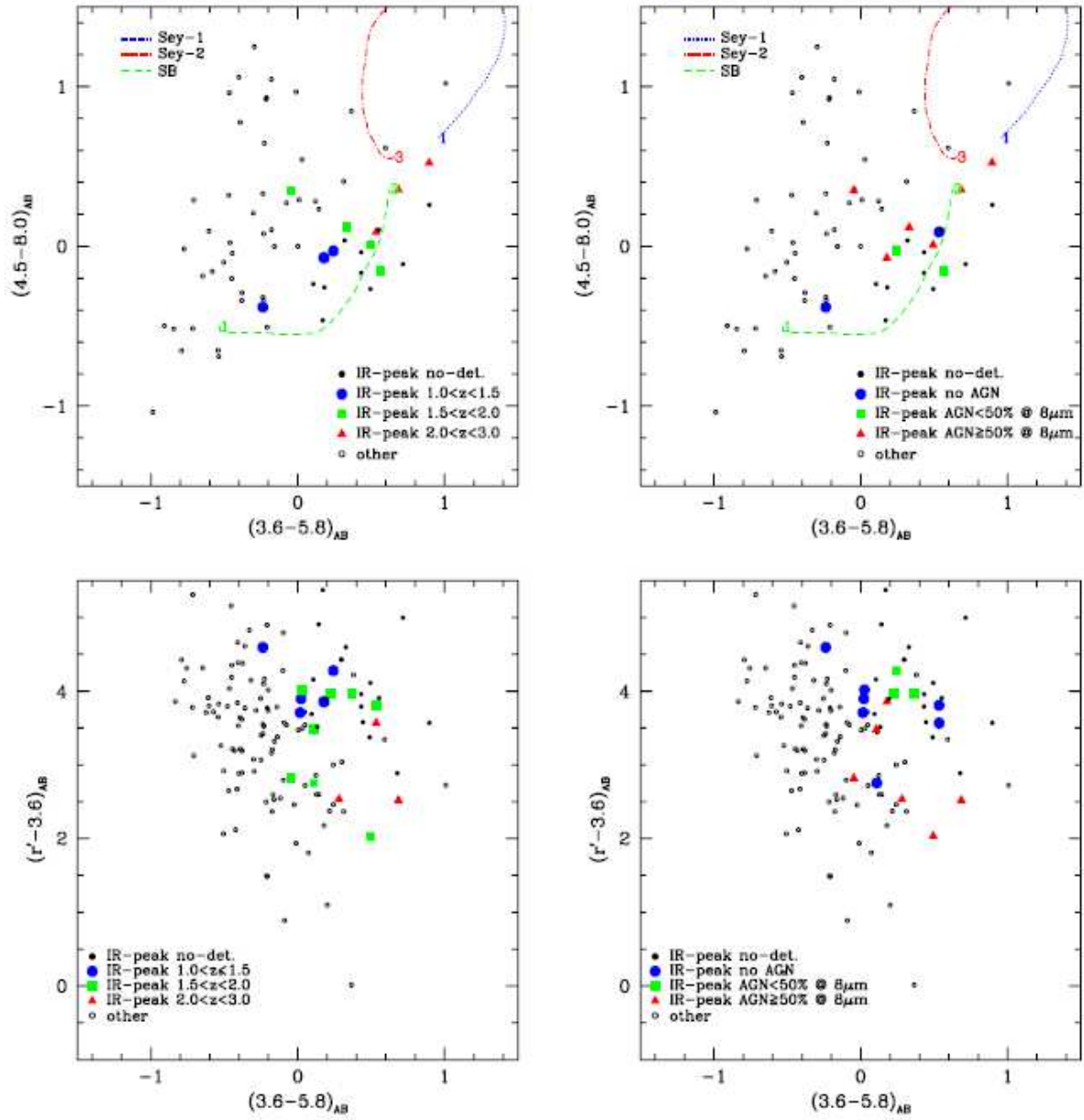}
\caption{Color-color plots of the targets with spectroscopic redshifts. 
The datapoints for IR-peak galaxies (detected in the required bands) are color-coded by
redshift ({\em left}) and AGN torus contribution at 8.0$\mu$m ({\em right}). 
Three template tracks from $z=1$ to $z=3$ are shown: a starburst (dashed lines, M82, Silva et al., \citeyear{silva1998}),
a seyfert-2 (dot-dash, IRAS 19254-7245, Berta et al., \citeyear{berta2003}) and a 
type-1 AGN (dotted, Mrk231, Fritz et al., \citeyear{fritz2006}).}
\label{fig:cc_coded}
\end{figure*}

Figure \ref{fig:cc_coded} shows the distribution of the IR-peak galaxies with 
a confirmed spectroscopic redshift, in optical-IRAC color space.
The points belonging to IR-peakers are color coded by redshift (left panels)
and by torus contribution at 8.0$\mu$m (right panels).
Redshift tracks for a Seyfert-1 \citep[Mrk231,][]{fritz2006}, a Seyfert-2
\citep[IRAS 19254-7245,][]{berta2003} and a starburst \citep[M82,][]{silva1998}
template are shown.

At comparable IRAC fluxes, 
the observed sources need to be increasingly optically blue with redshift
(bottom left panel), in order to be detected by Keck/LRIS, with reasonable 
exposure times. IRAC colors (top left) change in the same direction as the starburst track
(which is dominated by stars in the near-IR), but show a significant scatter, 
due to the torus contribution to the SEDs.
At the highest redshift end ($z\ge2.4$), the IRAC SED becomes flatter and resembles 
a power-law with a very diluted 1.6$\mu$m peak. The S(8.0)/S(4.5) flux ratio exceeds 
unity and sources transit to the AGN locus in the IRAC color space (triangles in the top left plot
of Fig. \ref{fig:cc_coded}).

As far as the torus contribution to the 8.0$\mu$m observed flux is concerned (right panels), 
the properties of the observed sources are not easily identified with the use
of broad band colors only.
No clear trend of the AGN fraction is seen in the IRAC color plot \citep[top right,][]{lacy2004}, 
for IR-peak galaxies, nor is segregation of different type of sources seen.
Apart from a couple of sources with power-law like SEDs (compare to. Fig. \ref{fig:ib_r_f24_lacy}), 
the IR-peak galaxies hosting an AGN tend to lie in the same locus of starburst galaxies 
in the IRAC color space.
The AGN contribution to IRAC SEDs has been identified:
\begin{enumerate}
\item through the presence of broad emission lines in their UV-optical spectra;
\item thanks to spectroscopic redshifts lower than expected, requiring a non negligible torus component
taking part in the near-IR emission of IR-peakers.
\end{enumerate}
In the optical, sources with 8.0$\mu$m torus fraction larger than 50\% seem to be
preferentially bluer than the others, as expected when the type-1 AGN component 
increasingly emerges at UV wavelengths (restframe). Type-2 AGNs show optical-IR
colors similar to starbursts, as their SEDs can even be fitted by stars alone.

The ``AGN IR-peak population'' contaminates the overall sample and is difficult 
to identify on the basis of broad band photometry alone. Extreme care should be taken 
in the analysis of sources in the fourth quadrant of the IRAC color plot,
the best way to break degeneracies and aliases being --- of course --- 
spectroscopic confirmation of their physical properties.

\section{Conclusions}

High redshift galaxies can be identified 
on the basis of of their restframe near-IR spectral energy distribution.
In this spectral domain, low-mass stars dominate galaxy emission, and 
produce a peak at 1.6$\mu$m (restframe), which is further enhanced by 
a minimum in the H$^-$ opacity in stellar atmospheres.
The stellar peak is detected in the IRAC channels 2 and 3 at redshifts between 1.5 and 3.0.

We have performed Keck/LRIS optical spectroscopy of high-$z$ ``IR-peak'' galaxies, 
selected in SWIRE northern fields. 
In order to be observable with Keck/LRIS, the sample was restricted to 
the optically brightest sources among the IR-peaker population.
A total of 35 such object were targeted,
and 16 have a spectroscopic confirmation, in the $z=1.3-2.8$ range.
Among these, six are 4.5$\mu$m-peakers, and the remaining
10 peak at 5.8$\mu$m. 

By combining the spectroscopic analysis and broad band SED fitting, 
we have extended our knowledge in the emission properties 
of IR-peak objects. The main results that have been described throughout
this paper are summarized below and in Tab. \ref{tab:summary}.

\begin{itemize}
\item The IRAC IR-peak galaxies turn out to lie in the redshift range $z=1.3-2.8$, 
broadly confirming expectations. 
Photometric and spectroscopic redshift are in better accordance for 4.5$\mu$m-peak objects
than for 5.8$\mu$m peakers, which turn out to be at slightly lower redshift than expected.
\item Low-redshift starburst interlopers represent a significant source of contamination of the 4.5$\mu$m-peak sample.
A bright 3.3$\mu$m PAH feature can significantly contribute to the IRAC channel 2
flux, for $z\sim 0.4$. Nevertheless, JHK data can break this aliasing,
sources with $(K_s-3.6)_{AB}\le0$ being at $z\le0.6$.
\item The optically-faintest IR-peakers that have very bright 24$\mu$m fluxes turn out to be
heavily extinguished starbursts, with $SFR>500$ M$_\odot$/yr and $A_V\ge2$ mag. No 
emission line are detected in these cases.
\item 69\% (11/16) of the IR-peakers with spectroscopic confirmation show AGN signatures in their spectra;
64\% (7/11) of these are broad-line type-1 objects, the remaining are type-2's.
The observed sample biased to the optically-brightest IR-peakers in 
the sky, likely favoring those which host optically-bright AGN.
\item On the basis of SED synthesis and spectral analysis, the 32\% (5/16) non-AGN sources are
powerful starbursts with SFR as high as $\sim500$ M$_\odot$/yr, 
stellar masses $M=1-6\times10^{11}$ M$_\odot$, and extinctions $A_V=1-2$ magnitudes.
The most active galaxies have specific SFRs fast enough to produce the 
bulk of the assembled stellar mass ($\sim10^{11}$ M$_\odot$) 
in one single burst of star formation.
\item All IR-peak broad-line AGN require a non-negligible contribution of torus warm dust 
to their IRAC SEDs; moreover, all sources that need a torus contribution to their mid-IR SED show 
AGN signatures in their UV-optical restframe spectra.
\item The AGN warm dust contribution to IRAC SEDs produces the apparent shift of the infrared peak
longward of 1.6$\mu$m (restframe), highlighted by the non perfect agreement
between photometric and spectroscopic redshifts.
In fact, the photometric estimate of redshift, based on stellar models, turned 
out to be frequently overestimated.
\item While IR-peakers follow a defined redshift track in the IRAC and optical-IR color space  (although with large scatter),
the AGN contamination of the sample can not be recognized on the basis of broad band colors only.
\item The AGN torus component is important in the IRAC domain, but it usually does not contribute 
significantly to longer wavelength mid-IR fluxes;
hence the mid-IR spectrum of these sources is dominated by PAH features, as confirmed by \cite{weedman2006}
IRS spectroscopy.
\end{itemize}

Finally, multi-object spectroscopy 
allowed us to include many other SWIRE sources on slits, for a total of 301 objects.  
Among these, 174 objects have a spectroscopic redshift; 150 targets with redshift have a SWIRE
counterpart. Our slits include 7 X-ray sources, 12 radio sources and 19 IRAC power-law galaxies.
On the basis of  spectral properties, we have identified 122 narrow-line emission galaxies,
39 turn out to be starbursts and 5 are type-2 AGNs. Seven targets have absorption lines only, 17 are 
broad-line type-1 AGNs and four are stars. Three high-redshift ($z\ge2.5$) QSOs 
complete the view of the targeted zoo.

\begin{acknowledgements}

We wish to thank the Keck night staff for their availability in 
producing calibration frames at the science targets ALT,AZ positions.
We are grateful to M. Meneghetti for tests on a supposed gravitational 
lens (which turned out not to be).
We thank A. Bressan and C. Chiosi for useful discussions on UV and near-IR
spectral features, S. Ciroi for suggestions on optical spectral classification, 
F. Di Mille and S. Siviero for elucidation on spectrographs and 
optical paths. 

The Spitzer Space Telescope
is operated by the Jet Propulsion Laboratory, California Institute of Technology, under
contract with NASA. 
SWIRE was supported by NASA through the SIRTF Legacy
Program under contract 1407 with the Jet Propulsion Laboratory.

\end{acknowledgements}


\bibliographystyle{aa}
\bibliography{6795bib}   



\begin{table*}[!ht]
\centering
\tiny 
\begin{tabular}{l c c c c c c c c l l c c }
\hline
\hline
ID & RA & DEC & r mag & S(3.6) & S(4.5) & S(5.8) & S(8.0) & S(24) & flag & Class & $z$ & $z$ \\
\# & $[$J2000$]$ & $[$J2000$]$ & $[$Vega$]$ & $[\mu$Jy$]$ & $[\mu$Jy$]$ & $[\mu$Jy$]$ & $[\mu$Jy$]$ & $[\mu$Jy$]$ &  & spec & spec & phot \\
\hline
EN1\_202261	&  242.330640	 &    54.676624       &  23.80   &    64.67  & 69.01  & 51.93  & 48.58  &   215.08    & P2       & ELG   &  1.339 &  1.190 \\
EN1\_205467	&  242.429490	 &    54.697529       &  23.40   &    36.37  & 55.60  & 83.10  & 70.62  &   488.24    & P3,X,pow & --    &  --    &  0.370 \\
EN1\_202260	&  242.401140	 &    54.636532       &  22.51   &    41.17  & 51.12  & 39.44  & 70.48  &   287.73    & P2,X     & BLAGN &  1.545 &  1.480 \\
EN1\_282078	&  244.260240	 &    55.522625       &  21.81   &    38.14  & 50.96  & 60.11  & 51.37  &   319.00    & P3       & BLAGN &  1.685 &  1.770 \\
EN1\_279954	&  244.184220	 &    55.511967       &  22.49   &    31.88  & 42.02  & 59.95  & 58.22  &   357.72    & P3,pow   & BLAGN &  2.409 &  2.770 \\
EN1\_341469	&  240.850890	 &    54.447605       &  24.07   &    103.34 & 119.90 & 120.89 & 78.24  &   992.89    & P3       & --    &  --    &  1.540 \\
EN1\_342445	&  240.908420	 &    54.440781       &  23.94   &    27.47  & 31.50  & 44.91  & --	&   213.89    & P3L      & ELG   &  1.917 &  1.850 \\
EN1\_340451	&  240.826570	 &    54.434467       &  23.32   &    38.87  & 54.19  & 63.95  & 58.84  &   181.78    & P3       & NLAGN &  2.866 &  2.320 \\
EN1\_339960	&  240.867170	 &    54.399624       &  23.75   &    50.34  & 62.56  & 62.94  & 60.88  &   248.89    & P3       & BLAGN &  1.475 &  1.590 \\
\hline
EN2\_275226	&  248.302110	 &    40.966358       &  23.74   &    38.23  & 53.61  & --     & 45.04  &   386.97    & P2       & BLAGN &  1.710 &  1.670 \\
EN2\_273717	&  248.339860	 &    40.951962       &  22.66   &    66.07  & 72.91  & 89.64  & 81.37  &   1121.32   & P3       & BLAGN &  1.800 &  3.050 \\
EN2\_269695	&  248.418080	 &    40.900318       &  23.17   &    42.83  & 48.45  & 57.50  & 50.11  &   768.10    & P3       & --    &  --    &  1.800 \\
EN2\_10334	&  250.438280	 &    40.763359       &  23.86   &    81.45  & 92.74  & 96.24  & 73.20  &   1073.62   & P3       & --    &  --    &  2.010 \\
EN2\_11091	&  250.464570	 &    40.791348       &  23.51   &    49.68  & 50.87  & 66.54  &  --	&   358.14    & P3L      & ELG   &  1.946 &  2.140 \\
EN2\_172324	&  248.621380	 &    41.059731       &  22.81   &    90.04  & 110.75 & 151.30 & 96.05  &   574.98    & P3       & NLAGN &  1.739 &  2.070 \\
EN2\_167372	&  248.650680	 &    40.979759       &  23.40   &    49.03  & 50.04  & --     &  --	&   --        & P2L      & ELG   &  1.445 &  2.120 \\
EN2\_166134	&  248.678600	 &    40.968796       &  22.87   &    67.08  & 68.06  & 44.93  &  --	&   216.85    & P2       & ELG   &  1.337 &  1.470 \\
EN2\_165986	&  248.712040	 &    40.980968       &  22.66   &    27.82  & 36.04  & 63.45  & 58.34  &   283.55    & P3,pow   & BLAGN &  2.163 &  0.430 \\
\hline
LH\_247598	&  163.005920	 &    57.852535       &  23.41   &    39.59  &  42.56 & 43.17  & --	&   241.24    & P3L      & --    &  --    &  1.130 \\
LH\_245973	&  163.022840	 &    57.795551       &  23.09   &    25.37  &  36.08 & 47.37  & --	&   605.48    & P3L      & --    &  --    &  2.840 \\
LH\_247451	&  163.070590	 &    57.811264       &  23.42   &    50.56  &  61.40 & 75.27  & 59.32  &   351.28    & P3       & --    &  --    &  1.790 \\
LH\_245782	&  163.066380	 &    57.765423       &  23.49   &    39.83  &  45.38 & 32.88  & --	&   --        & P2       & --    &  --    &  1.590 \\
LH\_247597	&  163.096080	 &    57.802212       &  23.24   &    51.16  &  60.40 & 76.15  & 51.76  &   871.90    & P3       & --    &  --    &  1.780 \\
LH\_571442	&  161.700320	 &    59.199207       &  24.31   &    21.28  &  30.67 & 35.41  & 33.76  &   --        & P3       & --    &  --    &  3.210 \\
LH\_572243	&  161.780960	 &    59.178944       &  22.94   &    25.94  &  30.26 & 28.69  & --	&   231.34    & P2,X     & NLAGN &  1.820 &  1.840 \\
LH\_575068	&  161.886430	 &    59.201855       &  24.44   &    30.45  &  34.50 & 40.03  & --	&   299.99    & P3L,R    & --    &  --    &  1.690 \\
LH\_572257	&  161.848860	 &    59.143932       &  22.91   &    24.85  &  33.69 & 34.50  & 31.46  &   229.24    & P3,X     & SB    &  0.249 &  2.170 \\
LH\_574364	&  161.909820	 &    59.169308       &  23.46   &    44.36  &  50.97 & 52.29  & 47.69  &   722.57    & P3,R     & NLAGN &  1.474 &  1.550 \\
LH\_125952	&  164.566760	 &    57.835228       &  24.46   &    34.68  &  38.04 & 46.91  & --	&   --        & P3L      & --    &  --    &  2.140 \\
LH\_126546	&  164.669680	 &    57.792194       &  23.18   &    36.65  &  46.40 & 57.55  & --	&   501.02    & P3L      & --    &  --    &  1.920 \\
LH\_128777	&  164.722290	 &    57.832966       &  24.37   &    54.40  &  71.86 & 105.16 & 64.84  &   819.83    & P3       & --    &  --    &  2.140 \\
LH\_577220	&  161.935620	 &    59.236961       &  23.69   &    27.75  &  34.55 & 41.80  & --	&   522.93    & P3L,R    & --    &  --    &  2.420 \\
LH\_576281	&  161.935760	 &    59.210655       &  23.99   &    35.83  &  44.53 & 39.46  & 35.84  &   416.95    & P2       & --    &  --    &  1.780 \\
LH\_574939	&  161.917590	 &    59.181812       &  23.47   &    45.26  &  50.26 & 54.08  & --	&   300.18    & P3L      & --    &  --    &  1.810 \\
LH\_577291	&  162.032840	 &    59.188950       &  23.81   &    40.68  &  49.33 & 64.18  & 38.59  &   407.16    & P3       & --    &  --    &  1.800 \\
\hline
\end{tabular}
\normalsize
\caption{Data for IR-peak sources included in slit, sorted by field: basic photometric information, 
photometric flag, spectroscopic classification and redshift are listed.
See Tab. \ref{tab:redshifts1} for more details on flags.}
\label{tab:IB_phot}
\end{table*}

\begin{table*}[!ht]
\centering
\tiny
\begin{tabular}{l | c c | c c c l | c}
\hline
\hline
ID & \multicolumn{2}{|c|}{Cont. $\lambda$ range $[$\AA$]$} &\multicolumn{4}{c|}{Detected lines} & $z$ \\									
SWIRE & blue & red & $\lambda$ $[$\AA$]$ & EW $[$\AA$]$ & FWHM $[$\AA$]$ & line & \\
\hline
LH\_247598	& 3500--5000	& 8800--7800	&	  --		  &	  --	  &	  --	  &	  --		  &	  --	  \\ 
LH\_245973	& --		& 6000--8500	&	  --		  &	  --	  &	  --	  &	  --		  &	  --	  \\ 
LH\_247451	& 4000--5000	& 5500--8300	&	  --		  &	  --	  &	  --	  &	  --		  &	  --	  \\ 
LH\_245782	& 4400--5500	& 6500--8000	&	  --		  &	  --	  &	  --	  &	  --		  &	  --	  \\ 
LH\_247597	& 3700--5300	& 5500--8000	&	  --		  &	  --	  &	  --	  &	  --		  &	  --	  \\ 
\hline
LH\_571442	& --		& 7100--9000	&	  --		  &	  --	  &	  --	  &	  --		  &	  --	  \\ 
LH\_572243	& 3300--5000	& 6100--8500	&	  3422.6	  &	  -116.2  &	$<$10.50  & Ly$\alpha$(1216)      &       1.815   \\ 
		&		& 		&	  4374.5	  &	  -36.3   &	11.20     & {\sc Civ]}(1549)      &       1.824   \\
		&		& 		&	  5383.4	  &	  --	  &	12.49     & {\sc Ciii]}(1909)     &       1.820   \\
LH\_575068	& 3500--4300	& 5600--8200	&	  --		  &	  --	  &	  --	  &	  --		  &	  --	  \\ 
LH\_572257	& 3300--5600	& 6200--8700	&	  4650.1	  &	  -63.3   &	$<$10.50  & {\sc [Oii]}(3727)     &       0.248   \\ 
		&		& 		&	  6255.9	  &	  -38.0   &	$<$10.50  & {\sc [Oiii]}(5007)    &       0.249   \\
		&		& 		&	  8193.4	  &	  -60.18  &	$<$10.50  &       H$\alpha$(6563) &       0.248   \\
LH\_574364	& 3400--5000	& 5700--8300	&	  6922.8	  &	  -114.0  &	  8.78    & {\sc Mgii}(2800)	  &	  1.474   \\				
\hline
EN1\_202261	& --		& 6200--9000	&	  8717.0	  &	  -39.3   &	$<$10.50  & {\sc [Oii]}(3727)	  &	  1.339   \\ 
EN1\_205467	& 3500--5300	& 5600--8300	&	  --		  &	  --	  &	  --	  &	  --		  &	  --	  \\ 
EN1\_202260	& 3000--5600	& 5700--9500	&	  3942.5	  &	  -449.8  &	  60.61   & {\sc Civ}(1549)	  &	  1.545   \\ 
		&		& 		&	  4169.3	  &	  -22.2   &	  21.62   & He{\sc ii}(1640)	  &	  1.541   \\
		&		& 		&	  4845.6	  &	  -111.8  &	  74.07   & {\sc Ciii]}1909	  &	  1.538   \\
		&		& 		&	  7119.7	  &	  -380.5  &	  154.24  & Mg{\sc ii}(2800)	  &	  1.545   \\
\hline
EN1\_282078	& 3100--5200	& 5500--8200	&	  3262.3	  &	  -23.7   &	  48.58   & Ly$\alpha$(1216)	  &	  1.683   \\ 
		&		& 		&	  3762.9	  &	  -62.7   &	  98.73   & Si{\sc iv,Oiv]}(1400) &	  1.688   \\
		&		& 		&	  4152.4	  &	  -118.9  &	  79.41   & {\sc Civ}(1549)	  &	  1.681   \\
		&		& 		&	  5673.6	  &	  -188.7  &	  239.87  &	  Fe{\sc ii}	  &	  --	  \\
		&		& 		&	  7517.2	  &	  -132.7  &	  133.29  & Mg{\sc ii}(2800)	  &	  1.687   \\
EN1\_279954	& 3200--5600	& 5700--8500	&	  4176.6	  &	  -242.3  &	  143.82  & Ly$\alpha$(1216)	  &	  2.435   \\ 
		&		&		&	  4157.7	  &	  -5.7    &	$<$10.50  & Ly$\alpha$(1216)        &       2.419   \\
		&		& 		&	  4173.9	  &	  -10.8   &	$<$10.50  & Ly$\alpha$(1216)        &       2.432   \\
		&		& 		&	  4778.3	  &	  -94.8   &	  108.90  & Si{\sc iv,Oiv]}(1400) &	  2.413   \\
		&		& 		&	  5280.7	  &	  -94.9   &	  68.99   & {\sc Civ}(1549)	  &	  2.409   \\
		&		& 		&	  5947.2	  &	  -98.7   &	  166.47  & {\sc Niii]}(1750)	  &	  2.398   \\
		&		& 		&	  6512.7	  &	  -198.2  &	  118.23  & {\sc Ciii]}(1909)	  &	  2.412   \\
\hline
EN2\_275226	& 3300--5700	& 5600--8500	&	  3377.6	  &	  -143.3  &	  17.29   & {\sc Nv}(1240)	  &	  1.724   \\ 
		&		& 		&	  4211.7	  &	  -294.3  &	  96.77   & {\sc Civ}(1549)	  &	  1.719   \\
		&		& 		&	  5153.7	  &	  -227.8  &	  130.78  & {\sc Ciii]}(1909)	  &	  1.700   \\
		&		& 		&	  7544.8	  &	  -126.9  &	  112.71  & Mg{\sc ii}(2800)	  &	  1.697   \\
EN2\_273717	& 3300--5600	& 5600--8600	&	  3478.8	  &	  -129    &	  $>$100  & {\sc Nv}(1240)	  &	  1.806   \\ 
		&		& 		&	  3949.7	  &	  -64	  &	  83.54   & Si{\sc iv,Oiv]}(1400) &	  1.821   \\
		&		& 		&	  4325.4	  & $<$-38.0	  &	  48.37   & {\sc Civ}(1549)	  &	  1.792   \\
		&		& 		&	  5347.8	  &	  -52.1   &	  98.64   & {\sc Ciii]}(1909)	  &	  1.801   \\
EN2\_269695	& 3200--5100	& 5600--8100	&	  --		  &	  --	  &	  --	  &	  --		  &	  --	  \\ 
\hline
\end{tabular}
\normalsize
\caption{Detected lines for IR-peak targets, sorted by mask (first night). For each source, we 
report the observed wavelength, equivalent width (EW), full width at half maximum (FWHM, corrected for 
the instrumental resolution) and identification of 
the detected spectral features. For each detected line, the spectroscopic 
redshift is computed (last column).}
\label{tab:spec_IB_1}
\end{table*}

\begin{table*}[!ht]
\centering
\tiny
\begin{tabular}{l | c c | c c c l | c}
\hline
\hline
ID & \multicolumn{2}{|c|}{Cont. $\lambda$ range $[$\AA$]$} &\multicolumn{4}{c|}{Detected lines} & $z$ \\									
SWIRE & blue & red & $\lambda$ $[$\AA$]$ & EW $[$\AA$]$ & FWHM $[$\AA$]$ & line & \\
\hline
LH\_125952	& --		& --		&	  --		  &	  --	  &	  --	  &	  --		  &	  --	  \\ 
LH\_126546	& 3500--5600	& 6000--8300	&	  --		  &	  --	  &	  --	  &	  --		  &	  --	  \\ 
LH\_128777	& --		& --		&	  --		  &	  --	  &	  --	  &	  --		  &	  --	  \\ 
\hline
LH\_575068	& --		& --		&	  --		  &	  --	  &	  --	  &	  --		  &	  --	  \\ 
LH\_577220	& --		& 5600--7600	&	  --		  &	  --	  &	  --	  &	  --		  &	  --	  \\ 
LH\_576281	& 3500--5600	& 5600--8600	&	  --		  &	  --	  &	  --	  & --  		  &	  --	  \\ 
LH\_574939	& 3400--5000	& --		&	  --		  &	  --	  &	  --	  &	  --		  &	  --	  \\ 
LH\_574364	& --		& 6000--8500	&	  6521.87	  &	  -39.66  &	  $<$10.50&	  Fe		  &	  --	  \\ 
		&		& 		&	  6923.38	  &	  -100.2  &	  10.95   & Mg{\sc ii}(2800)	  &	  1.474   \\
		&		& 		&	  9209.20	  &	  -125.6  &	  7.34    & {\sc [Oii]}(3727)	  &	  1.471   \\
LH\_577291	& --		& 5700--7500	&	  --		  &	  --	  &	  --	  &	  --		  &	  --	  \\ 
\hline
EN1\_341469	& --		& 5700--8500	&	  --		  &	  --	  &	  --	  &	  --		  &	  --	  \\ 
EN1\_342445	& 3300--5500	& 5600--8500	&	  3545.82	  &	  -177    &	  6.31    & Ly$\alpha$(1216)	  &	  1.916   \\ 
		&		& 		&	  4518.67	  &	  -47.75  &	  13.76   & {\sc Civ}(1549)	  &	  1.917   \\
EN1\_340451	& --		& --		&	  4702.81	  &	  -141.7  &	  8.92    & Ly$\alpha$(1216)	  &	  2.867   \\ 
		&		& 		&	  6343.29	  &	  -21.04  &	  14.87   & He{\sc ii}(1640)	  &	  2.868   \\
		&		& 		&	  7373.11  	  &	  --	  &	  --	  & {\sc Ciii]}(1909)	  &	  2.862   \\
EN1\_339960	& 3600--5600	& 5700--8500	&	  3834.78	  &	  -101.9  &	  27.66   & {\sc Civ}(1549)	  &	  1.476   \\ 
		&		& 		&	  4720.29	  &	  -47.48  &	  28.67   & {\sc Ciii]}(1909)	  &	  1.473   \\
\hline
EN2\_10334	& 4000--5600	& 6000--8500	&	  --		  &	  --	  &	  --	  &	  --		  &	  --	  \\ 
EN2\_11091	& 4000--5600	& 5700--8200	&	  3581.76	  &	  $<$60   &	  5.75    & Ly$\alpha$(1216)	  &	  1.946   \\ 
\hline
EN2\_172324	& 3100--5700	& 5700--8500	&	  3333.35	  &	  -68.92  &	  7.36    & Ly$\alpha$(1216)	  &	  1.741   \\ 
		&		& 		&	  4250.45	  &	  -41.41  &	  25.17   & {\sc Civ]}(1549)	  &	  1.744   \\
		&		& 		&	  4484.25	  &	  -24.33  &	  45.54   & He{\sc ii}(1640)	  &	  1.734   \\
		&		& 		&	  5223.39	  &	  -21.81  &	  21.37   & {\sc Ciii]}(1909)	  &	  1.736   \\
EN2\_167372	& 3200--5600	& 6000--9200	&	  9112.03	  &	  -35.71  &	  19.06   & {\sc [Oii]}(3727)	  &	  1.445   \\ 
EN2\_166134	& 3200--5700	& 5700--9000	&	  8711.26	  &	  -31.09  &	  10.23   & {\sc [Oii]}(3727)	  &	  1.337   \\ 
EN2\_165986	& 3200--5700	& 5600--8700	&	  3845.90	  &	  --	  &	  $<$10.50& Ly$\alpha$(1216)	  &	  2.163   \\ 
		&		& 		&	  3872.52	  &	  -499.2  &	  146.83  & Ly$\alpha$(1216)	  &	  2.185   \\ 
		&		& 		&	  4433.21	  &	  -51.1   &	  119.74  & Si{\sc iv,Oiv]}(1400) &	  2.167   \\
		&		& 		&	  4897.26	  &	  -193.6  &	  118.24  & {\sc Civ}(1549)	  &	  2.162   \\ 
		&		& 		&	  6026.88	  &	  -240.6  &	  225.46  & {\sc Ciii]}(1909)	  &	  2.157   \\ 
\hline
\end{tabular}
\normalsize
\caption{Detected lines for IR-peak targets, sorted by mask (second night).}
\label{tab:spec_IB_2}
\end{table*}

\begin{table*}[!ht]
\centering
\tiny
\begin{tabular}{l c c c c c c l l}
\hline
\hline
ID$^\ddagger$ 	& RA 	& DEC	& \multicolumn{2}{c}{Det. Lines} & $z$ & $z$ &Note$^\ast$ & Class$^\dagger$\\
SWIRE	& J2000	& J2000	& em. & abs. & spec & phot & & spec\\
\hline
LH\_247176	&	162.9559600	&	57.8670580	&	0	&	0	&	--		&	0.720	&	& -- \\
LH\_245141	&	162.9474900	&	57.8128240	&	1	&	0	&	1.017		&	0.890	&	& ELG \\
LH\_245572	&	162.9637900	&	57.8163990	&	1	&	0	&	0.935		&	0.890	&	& ELG \\
LH\_244996	&	162.9585700	&	57.8022500	&	0	&	0	&	--		&	1.240	&	& -- \\
LH\_244783	&	162.9599600	&	57.7954600	&	0	&	0	&	--		&	0.590	&	& -- \\
LH\_247598	&	163.0059200	&	57.8525350	&	0	&	0	&	--		&	1.130	& P3L	& -- \\
LH\_246388	&	163.0071900	&	57.8170200	&	2	&	0	&	0.700		&	1.030	&	& ELG \\
LH\_247704	&	163.0369100	&	57.8389550	&	1	&	0	&	0.419		&	0.500	&	& ELG \\
LH\_246801	&	163.0273900	&	57.8167570	&	0	&	0	&	--		&	1.420	&	& -- \\
LH\_245973	&	163.0228400	&	57.7955510	&	0	&	0	&	--		&	2.840	& P3L	& -- \\
LH\_248065	&	163.0620100	&	57.8371960	&	0	&	0	&	--		&	1.000	&	& -- \\
LH\_246098	&	163.0397000	&	57.7900430	&	1	&	0	&	0.927		&	0.400	&	& ELG \\
LH\_246660	&	163.0540000	&	57.7978250	&	0	&	0	&	--		&	2.780	&	& -- \\
LH\_247451	&	163.0705900	&	57.8112640	&	0	&	0	&	--		&	1.790	& P3	& -- \\
LH\_245683	&	163.0602400	&	57.7662620	&	0	&	0	&	--		&	1.130	&	& -- \\
LH\_245782	&	163.0663800	&	57.7654230	&	0	&	0	&	--		&	1.590	& P2	& -- \\
LH\_247597	&	163.0960800	&	57.8022120	&	0	&	0	&	--		&	1.780	& P3	& -- \\
LH\_247508	&	163.1037600	&	57.7950170	&	0	&	0	&	--		&	0.980	&	& -- \\
LH\_247717	&	163.1172000	&	57.7946130	&	1	&	0	&	1.026		&	1.000	&	& ELG \\
LH\_248867	&	163.1425500	&	57.8190800	&	0	&	0	&	--		&	1.130	&	& -- \\
LH\_246777	&	163.1230900	&	57.7627410	&	0	&	0	&	--		&	4.410	&	& -- \\
\hline
LH\_574421	&	161.7675900	&	59.2445530	&	0	&	0	&	--		&	0.960	&       & -- \\
LH\_571442	&	161.7003200	&	59.1992070	&	0	&	0	&	--		&	3.210	& P3    & -- \\
LH\_574156	&	161.7721700	&	59.2355310	&	0	&	0	&	--		&	1.000	&       & -- \\
LH\_575325	&	161.8065900	&	59.2504010	&	5	&	0	&	0.355		&	0.310	& X,R,pow & BLAGN   \\
LH\_574146	&	161.7805300	&	59.2307780	&	1	&	3	&	0.850		&	0.790	&       & SB \\
LH\_571907	&	161.7359900	&	59.1932950	&	5	&	0	&	0.504		&	1.400	& R     & SB \\
LH\_571884 (s)	&	161.7322700	&	59.1946200	&	0	&	0	&	--		&	0.350	&       & -- \\ 
LH\_574490	&	161.8032800	&	59.2278290	&	0	&	0	&	--		&	0.880	&       & -- \\
LH\_574646	&	161.8123900	&	59.2275850	&	1	&	0	&	0.712		&	0.570	& R     & ELG \\
LH\_572854	&	161.7723100	&	59.2003100	&	1	&	0	&	0.425		&	0.590	&       & ELG \\
LH\_572196	&	161.7604200	&	59.1882060	&	2	&	0	&	0.448		&	0.810	&       & ELG \\
LH\_572282	&	161.7674300	&	59.1873470	&	0	&	0	&	--		&	0.390	&       & -- \\
LH\_573252	&	161.7947200	&	59.1993520	&	1	&	0	&	0.826		&	0.890	& R     & ELG \\
LH\_572243	&	161.7809600	&	59.1789440	&	3	&	0	&	1.820		&	1.840	& P2,X  & NLAGN \\
LH\_573752	&	161.8214400	&	59.1984940	&	1	&	0	&	0.871		&	1.060	&       & ELG \\
LH\_572227	&	161.7896100	&	59.1740110	&	0	&	0	&	--		&	0.780	&       & -- \\
LH\_575068	&	161.8864300	&	59.2018550	&	0	&	0	&	--		&	1.690	& P3L,R & -- \\
LH\_574218	&	161.8758100	&	59.1835670	&	1	&	0	&	1.169		&	1.080	&       & ELG \\
LH\_572289	&	161.8328600	&	59.1532900	&	1	&	0	&	0.992		&	1.120	& R     & ELG \\
LH\_573719	&	161.8779400	&	59.1690940	&	0	&	0	&	--		&	1.240	&       & -- \\
LH\_572257	&	161.8488600	&	59.1439320	&	3	&	1	&	0.249		&	2.170	& P3,X  & SB \\
LH\_571738	&	161.8405300	&	59.1352350	&	0	&	0	&	--		&	1.210	&       & -- \\
LH\_574364	&	161.9098200	&	59.1693080	&	1	&	0	&	1.474		&	1.550	& P3,R  & NLAGN \\
LH\_572820	&	161.8769100	&	59.1450000	&	6	&	0	&	0.305		&	0.590	&       & SB \\
LH\_574495	&	161.9221600	&	59.1664850	&	0	&	0	&	--		&	1.200	&       & -- \\
LH\_572108	&	161.8726800	&	59.1277350	&	1	&	0	&	1.019		&	0.880	& R     & ELG \\
LH\_572670	&	161.8921700	&	59.1328200	&	0	&	0	&	--		&	1.070	& R     & -- \\
\hline
EN1\_203038	&	242.2878400	&	54.7196460	&	0	&	0	&	--		&	1.040	&       & -- \\
EN1\_204462	&	242.3317600	&	54.7288320	&	0	&	0	&	--		&	1.310	&       & -- \\
EN1\_206332	&	242.4030800	&	54.7331850	&	1	&	6	&	0.754		&	0.490	&       & ELG \\
EN1\_205851	&	242.3945300	&	54.7264710	&	0	&	0	&	--		&	1.100	&       & -- \\
EN1\_203688	&	242.3373700	&	54.7072750	&	1	&	0	&	0.737		&	1.010	&       & ELG \\
EN1\_204660	&	242.3713100	&	54.7115440	&	4	&	7	&	0.630		&	0.550	&       & SB \\
EN1\_202683	&	242.3188800	&	54.6936190	&	5	&	0	&	0.498		&	0.550	&       & SB \\
EN1\_202789	&	242.3312800	&	54.6889000	&	0	&	0	&	--		&	0.910	&       & -- \\
EN1\_204678	&	242.3887200	&	54.7020150	&	2	&	0	&	0.469		&	0.440	&       & ELG \\
EN1\_204710	&	242.3942400	&	54.6996460	&	0	&	3	&	0.874		&	1.100	&       & ALG \\
EN1\_204752 (s)	&	242.3934800	&	54.7009500	&	4	&	0	&	0.469		&	0.200	&       & SB \\ 
EN1\_202261	&	242.3306400	&	54.6766240	&	1	&	0	&	1.339		&	1.190	& P2    & ELG \\
EN1\_205467	&	242.4294900	&	54.6975290	&	0	&	0	&	--		&	0.370	& P3,X,pow & -- \\
EN1\_203073	&	242.3640000	&	54.6769600	&	0	&	0	&	--		&	0.710	&       & -- \\
EN1\_203962	&	242.4050100	&	54.6756210	&	8	&	0	&	0.874		&	0.860	& X     & NLAGN\\
EN1\_202756	&	242.3763400	&	54.6624410	&	6	&	0	&	3.005		&	2.720	&       & BLAGN\\
EN1\_204120	&	242.4209400	&	54.6700710	&	3	&	0	&	1.475		&	0.350	& X,pow & BLAGN\\
EN1\_202023	&	242.3639100	&	54.6516880	&	1	&	3	&	0.898		&	1.010	&       & SB \\
EN1\_205340	&	242.4640000	&	54.6744960	&	1	&	0	&	0.905		&	0.910	&       & ELG \\
EN1\_202047	&	242.3729400	&	54.6469380	&	1	&	5	&	0.661		&	0.550	&       & SB \\
EN1\_204124	&	242.4431600	&	54.6573910	&	1	&	0	&	0.901		&	1.070	&       & ELG \\
EN1\_202260	&	242.4011400	&	54.6365320	&	4	&	0	&	1.545		&	1.480	& P2,X  & BLAGN\\
EN1\_205034	&	242.4857800	&	54.6545030	&	1	&	0	&	0.876		&	0.910	&       & ELG \\
EN1\_203525	&	242.4547000	&	54.6364940	&	2	&	0	&	0.265		&	0.210	&       & ELG \\
EN1\_201165	&	242.3914200	&	54.6144640	&	1	&	2	&	0.762		&	0.220	& X,pow & SB \\
\hline
\multicolumn{9}{l}{$^\ddagger$: (s) indicates serendipitous sources; }\\
\multicolumn{9}{l}{$^\ast$: pow $=$ IRAC power-law SED; P3 $=$ 5.8$\mu$m peaker; P2 $=$ 4.5$\mu$m peaker; }\\
\multicolumn{9}{l}{\ \ \ L $=$ upper limit at 8.0$\mu$m; R $=$ radio source; X $=$ X-ray source.}\\
\multicolumn{9}{l}{$^\dagger$: ELG $=$ emission lines; SB $=$ starburst diagnostics; BLAGN $=$ broad line AGN; NLAGN $=$}\\
\multicolumn{9}{l}{\ \ \ narrow-line AGN; star $=$ star; ALG $=$ absorption lines only; nc $=$ no continuum detected}\\
\multicolumn{9}{l}{\ \ \ (but emission lines yes).}\\
\end{tabular}
\normalsize
\caption{Results for first night. For each target, the number of detected 
absorption and emission lines is reported, as well as a photometric and spectroscopic
(see footnotes) classification. The measured spectroscopic redshift is in Column 6, 
while column 7 lists the photometric estimate of $z$, as obtained with the
Hyper-z code \citep{bolzonella2000}.}
\label{tab:redshifts1}
\end{table*}

\begin{table*}[!ht]
\centering
\tiny
\begin{tabular}{l c c c c c c l l}
\hline
\hline
ID$^\ddagger$ 	& RA 	& DEC	& \multicolumn{2}{c}{Det. Lines} & $z$ & $z$& Note$^\ast$ & Class$^\dagger$\\
SWIRE	& J2000	& J2000	& em. & abs. & spec & phot & & spec\\
\hline
EN1\_283675	&	244.1841300	&	55.6057700	&	1	&	0	&	1.138		&	0.870	&       & ELG \\
EN1\_284886	&	244.2396200	&	55.6035190	&	0	&	2	&	0.800		&	0.810	&       & ALG\\
EN1\_285168	&	244.2629100	&	55.5971790	&	0	&	2	&	0.798		&	0.920	&       & ALG\\
EN1\_282051	&	244.1426400	&	55.5912250	&	4	&	0	&	3.100		&	0.210	& pow	& BLAGN\\
EN1\_281866	&	244.1489000	&	55.5828820	&	1	&	0	&	0.732		&	0.900	&       & ELG \\
EN1\_282887	&	244.1957600	&	55.5805400	&	0	&	0	&	--		&	1.370	&       & -- \\
EN1\_283904	&	244.2396100	&	55.5787850	&	0	&	0	&	--		&	0.980	&       & -- \\
EN1\_283549	&	244.2381300	&	55.5712430	&	1	&	0	&	1.119		&	1.100	&       & ELG \\
EN1\_282211	&	244.1892700	&	55.5676650	&	1	&	0	&	1.103		&	1.120	&       & ELG \\
EN1\_282263 (s)	&	244.2012600	&	55.5621000	&	1	&	0	&	0.808		&	0.720	&       & ELG \\ 
EN1\_282169	&	244.2009400	&	55.5596310	&	4	&	0	&	0.461		&	0.970	&       & ELG \\
EN1\_281970 (s)	&	244.2005200	&	55.5551800	&	1	&	0	&	0.807		&	0.580	&       & ELG \\ 
EN1\_282870	&	244.2621500	&	55.5409550	&	0	&	0	&	--		&	0.910	&       & -- \\
EN1\_281341	&	244.2031100	&	55.5369150	&	1	&	0	&	0.868		&	0.880	&       & ELG \\
EN1\_279813	&	244.1436800	&	55.5324780	&	2	&	4	&	0.720		&	0.730	&       & SB \\
EN1\_279609	&	244.1402700	&	55.5297660	&	1	&	0	&	1.165		&	0.910	&       & ELG \\
EN1\_281524	&	244.2280600	&	55.5273400	&	0	&	0	&	--		&	1.370	&       & -- \\
EN1\_282078	&	244.2602400	&	55.5226250	&	4	&	0	&	1.685		&	1.770	& P3    & BLAGN\\
EN1\_280187	&	244.1820500	&	55.5190510	&	1	&	0	&	0.891		&	0.980	&       & ELG \\
EN1\_281722	&	244.2551400	&	55.5166890	&	1	&	0	&	1.211		&	1.770	&       & ELG \\
EN1\_279954	&	244.1842200	&	55.5119670	&	5	&	0	&	2.409		&	2.770	& P3,pow& BLAGN\\
EN1\_279984 (s)	&	244.1844600	&	55.5128000	&	2	&	0	&	0.718		&	0.700	&       & ELG \\ 
EN1\_281198	&	244.2435300	&	55.5091170	&	0	&	0	&	--		&	1.070	&       & -- \\
EN1\_280713	&	244.2269600	&	55.5062330	&	1	&	0	&	1.211		&	1.420	&       & ELG \\
EN1\_280065 (s)	&	244.2111400	&	55.4988300	&	7	&	5	&	0.102		&	0.100	&       & SB \\  
EN1\_279938	&	244.2106500	&	55.4962500	&	1	&	5	&	0.903		&	1.080	&       & SB \\
EN1\_279170	&	244.1830700	&	55.4936370	&	0	&	0	&	--		&	1.320	&       & -- \\
EN1\_280103	&	244.2275200	&	55.4899600	&	1	&	0	&	1.171		&	0.980	&       & ELG \\
\hline	
EN2\_275885	&	248.2923000	&	40.9753460	&	0	&	0	&	--		&	0.600	&       & -- \\
EN2\_275543	&	248.2659500	&	40.9568060	&	6	&	7	&	0.392		&	0.210	&       & SB \\
EN2\_275226	&	248.3021100	&	40.9663580	&	4	&	0	&	1.710		&	1.670	& P2    & BLAGN\\
EN2\_274748 (s)	&	248.3105500	&	40.9599700	&	--	&	--	&	0.000		&	--	&       & star\\ 
EN2\_274821	&	248.3091000	&	40.9605980	&	1	&	0	&	0.909		&	0.970	&       & ELG \\
EN2\_274735	&	248.3425600	&	40.9739490	&	5	&	0	&	2.605		&	2.690	& pow   & BLAGN\\
EN2\_273908	&	248.2948200	&	40.9358180	&	0	&	0	&	--		&	2.840	&       & -- \\
EN2\_273814	&	248.3073100	&	40.9395140	&	0	&	0	&	--		&	1.610	&       & -- \\
EN2\_273717	&	248.3398600	&	40.9519620	&	3	&	0	&	1.800		&	3.050	& P3    & BLAGN\\
EN2\_273639	&	248.3609200	&	40.9597550	&	1	&	6	&	0.704		&	0.720	&       & SB \\
EN2\_273300	&	248.3362700	&	40.9415890	&	1	&	0	&	0.987		&	1.710	& pow   & ELG \\
EN2\_272775	&	248.3101800	&	40.9183240	&	5	&	4	&	0.384		&	0.170	&       & SB \\
EN2\_272288	&	248.3251800	&	40.9139370	&	1	&	5	&	0.787		&	0.700	&       & SB \\
EN2\_272228	&	248.3427700	&	40.9206160	&	0	&	0	&	--		&	1.550	&       & -- \\
EN2\_271541	&	248.3325800	&	40.9015120	&	1	&	0	&	1.156		&	1.100	&       & ELG \\
EN2\_271345 	&	248.3609800	&	40.9107890	&	0	&	0	&	 --		&	0.410	&       & -- \\
EN2\_271260 	&	248.3783700	&	40.9168550	&	4	&	0	&	 1.820  	&	0.500	& pow   & BLAGN\\
EN2\_270481	&	248.3485400	&	40.8865470	&	1	&	0	&	1.186		&	1.220	&       & ELG \\
EN2\_270623	&	248.4030900	&	40.9140590	&	0	&	0	&	--		&	1.470	&       & -- \\
EN2\_270327	&	248.3833800	&	40.8986470	&	1	&	0	&	1.061		&	1.050	&       & ELG \\
EN2\_270081	&	248.3608400	&	40.8828390	&	1	&	0	&	1.187		&	1.100	&       & ELG \\
EN2\_270049	&	248.3937700	&	40.8965800	&	0	&	0	&	--		&	1.030	&       & -- \\
EN2\_269695	&	248.4180800	&	40.9003180	&	0	&	0	&	--		&	1.800	& P3    & -- \\
EN2\_269090	&	248.3551200	&	40.8594930	&	1	&	0	&	1.062		&	0.720	&       & ELG \\
\hline
\multicolumn{9}{l}{$^\ddagger$: (s) indicates serendipitous sources; }\\
\multicolumn{9}{l}{$^\ast$: pow $=$ IRAC power-law SED; P3 $=$ 5.8$\mu$m peaker; P2 $=$ 4.5$\mu$m peaker; }\\
\multicolumn{9}{l}{\ \ \ L $=$ upper limit at 8.0$\mu$m; R $=$ radio source; X $=$ X-ray source.}\\
\multicolumn{9}{l}{$^\dagger$: ELG $=$ emission lines; SB $=$ starburst diagnostics; BLAGN $=$ broad line AGN; NLAGN $=$}\\
\multicolumn{9}{l}{\ \ \ narrow-line AGN; star $=$ star; ALG $=$ absorption lines only; nc $=$ no continuum detected}\\
\multicolumn{9}{l}{\ \ \ (but emission lines yes).}\\
\end{tabular}
\normalsize
\begin{flushleft}
{\bf Table \ref{tab:redshifts1}.} Continued.
\end{flushleft}
\end{table*}

\begin{table*}[!ht]
\centering
\tiny
\begin{tabular}{l c c c c c c l l}
\hline
\hline
ID$^\ddagger$ 	& RA 	& DEC	& \multicolumn{2}{c}{Det. Lines} & $z$ & $z$ & Note$^\ast$ & Class$^\dagger$\\
SWIRE	& J2000	& J2000	& em. & abs. & spec & phot & & spec\\
\hline
LH\_126754	&	164.5717500	&	57.8558810	&	0	&	0	&	--		&	2.950	&       & -- \\
LH\_125952	&	164.5667600	&	57.8352280	&	0	&	0	&	--		&	2.140	& P3L   & -- \\
LH\_125107	&	164.5657700	&	57.8102190	&	1	&	0	&	0.974		&	1.090	&       & ELG \\
LH\_126739	&	164.5884700	&	57.8456920	&	0	&	0	&	--		&	1.040	&       & -- \\
LH\_125753	&	164.5841200	&	57.8190310	&	0	&	0	&	--		&	0.910	&       & -- \\
LH\_124994	&	164.5825700	&	57.7961500	&	0	&	0	&	--		&	1.160	&       & -- \\
LH\_125213	&	164.5899700	&	57.7990530	&	5	&	0	&	0.477		&	0.210	&       & SB \\
LH\_125861	&	164.6013600	&	57.8122560	&	0	&	0	&	--		&	0.570	&       & -- \\
LH\_128120	&	164.6321100	&	57.8646160	&	0	&	0	&	--		&	0.880	&       & -- \\
LH\_127702	&	164.6326800	&	57.8511200	&	0	&	0	&	--		&	1.180	&       & -- \\
LH\_127862	&	164.6456500	&	57.8481830	&	0	&	0	&	--		&	1.240	&       & -- \\
LH\_128064	&	164.6524000	&	57.8508000	&	1	&	0	&	2.233		&	1.500	& pow   & ELG \\
LH\_127481	&	164.6514900	&	57.8335190	&	1	&	0	&	0.854		&	0.740	&       & ELG \\
LH\_127839	&	164.6627500	&	57.8373490	&	0	&	0	&	--		&	0.950	&       & -- \\
LH\_128218	&	164.6739800	&	57.8436010	&	1	&	0	&	0.932		&	1.370	&       & ELG \\
LH\_126546	&	164.6696800	&	57.7921940	&	0	&	0	&	--		&	1.920	& P3L   & -- \\
LH\_127562	&	164.6873800	&	57.8152280	&	1	&	0	&	1.200		&	1.320	&       & ELG \\
LH\_128115	&	164.6974900	&	57.8262560	&	1	&	0	&	0.741		&	0.860	&       & ELG \\
LH\_127706	&	164.7046700	&	57.8094410	&	1	&	0	&	0.965		&	0.970	&       & ELG \\
LH\_128777	&	164.7222900	&	57.8329660	&	0	&	0	&	--		&	2.140	& P3    & -- \\
LH\_129098	&	164.7355200	&	57.8357200	&	0	&	0	&	--		&	1.290	&       & -- \\
LH\_129336	&	164.7433200	&	57.8387790	&	0	&	0	&	--		&	1.190	&       & -- \\
LH\_129463	&	164.7498500	&	57.8388670	&	0	&	0	&	--		&	1.000	&       & -- \\
LH\_129057	&	164.7492200	&	57.8263590	&	1	&	0	&	1.025		&	0.970	&       & ELG \\
LH\_128839	&	164.7504400	&	57.8182180	&	3	&	0	&	0.441		&	0.230	&       & ELG \\
LH\_127252	&	164.7381900	&	57.7757530	&	0	&	0	&	--		&	0.100	&       & -- \\
LH\_129519	&	164.7719900	&	57.8275340	&	0	&	0	&	--		&	0.980	&       & -- \\
LH\_129657	&	164.7800300	&	57.8278500	&	1	&	0	&	0.903		&	0.940	&       & ELG \\
LH\_129341	&	164.7840900	&	57.8150900	&	0	&	0	&	--		&	1.090	&       & -- \\
\hline
LH\_573671	&	161.8165100	&	59.1993370	&	0	&	0	&	--		&	1.220	&       & -- \\
LH\_576445	&	161.8730800	&	59.2471200	&	1	&	0	&	2.024		&	2.150	& pow   & ELG \\
LH\_576161	&	161.8712900	&	59.2399900	&	1	&	0	&	0.988		&	0.970	&       & ELG \\
LH\_573414	&	161.8264300	&	59.1871410	&	1	&	4	&	0.749		&	0.980	& R     & SB \\
LH\_573584	&	161.8378800	&	59.1861690	&	0	&	0	&	--		&	0.290	&       & -- \\
LH\_576452	&	161.8964700	&	59.2355460	&	0	&	0	&	--		&	2.990	&       & -- \\
LH\_573395	&	161.8467400	&	59.1762120	&	0	&	0	&	--		&	1.000	&       & -- \\
LH\_577039	&	161.9206100	&	59.2402080	&	0	&	0	&	--		&	0.990	&       & -- \\
LH\_575068	&	161.8864300	&	59.2018550	&	0	&	0	&	--		&	1.690	& P3L,R & -- \\
LH\_577220	&	161.9356200	&	59.2369610	&	0	&	0	&	--		&	2.420	& P3L,R & -- \\
LH\_574211	&	161.8810300	&	59.1807750	&	0	&	0	&	--		&	0.900	&       & -- \\
LH\_577256	&	161.9487800	&	59.2319870	&	1	&	0	&	1.127		&	1.200	&       & ELG \\
LH\_576281	&	161.9357600	&	59.2106550	&	1	&	0	&	--		&	1.780	& P2    & ELG \\
LH\_574939	&	161.9175900	&	59.1818120	&	0	&	0	&	--		&	1.810	& P3L   & -- \\
LH\_574364	&	161.9098200	&	59.1693080	&	3	&	0	&	1.474		&	1.550	& P3,R  & NLAGN \\
LH\_576769	&	161.9698800	&	59.2069210	&	3	&	0	&	0.659		&	0.260	& R,pow & ELG \\
LH\_574561	&	161.9313400	&	59.1634520	&	0	&	0	&	--		&	1.130	&       & -- \\
LH\_577480	&	161.9990700	&	59.2114450	&	1	&	0	&	0.581		&	0.800	&       & ELG \\
LH\_574430 (s)	&	161.9459800	&	59.1525500	&	0	&	0	&	0.000		&	--	&       & star \\ 
LH\_574491	&	161.9507600	&	59.1516720	&	0	&	0	&	--		&	0.500	& pow   & -- \\
LH\_575939	&	161.9877600	&	59.1729700	&	2	&	0	&	0.128		&	0.180	&       & ELG \\
LH\_576851	&	162.0119600	&	59.1873510	&	0	&	0	&	--		&	1.150	&       & -- \\
LH\_576029 (s)	&	162.0022400	&	59.1681800	&	1	&	0	&	1.246		&	0.650	&       & ELG \\ 
LH\_575976	&	161.9977000	&	59.1693080	&	1	&	0	&	1.023		&	1.090	&       & ELG \\
LH\_577291	&	162.0328400	&	59.1889500	&	1	&	0	&	--		&	1.800	& P3    & -- \\
LH\_577043	&	162.0330000	&	59.1823390	&	0	&	0	&	--		&	0.970	&       & -- \\
\hline	
EN1\_344017	&	240.9222700	&	54.4750630	&	0	&	0	&	--		&	0.510	&       & -- \\
EN1\_341502	&	240.8081500	&	54.4725880	&	7	&	0	&	2.317		&	2.910	& pow   & BLAGN\\
EN1\_341999	&	240.8352800	&	54.4703790	&	0	&	0	&	--		&	1.220	&       & -- \\
EN1\_343159	&	240.8965600	&	54.4654690	&	0	&	0	&	--		&	0.880	&       & -- \\
EN1\_343609	&	240.9236000	&	54.4629630	&	0	&	4	&	0.547		&	0.390	&       & ALG\\
EN1\_342460	&	240.8772700	&	54.4583630	&	1	&	0	&	0.757		&	0.490	&       & ELG \\
EN1\_340789	&	240.8124200	&	54.4507180	&	1	&	0	&	1.092		&	0.480	&       & ELG \\
EN1\_341469	&	240.8508900	&	54.4476050	&	0	&	0	&	--		&	1.540	& P3    & -- \\
EN1\_343359	&	240.9443800	&	54.4448320	&	1	&	2	&	0.815		&	0.680	&       & SB \\
EN1\_342445	&	240.9084200	&	54.4407810	&	2	&	0	&	1.917		&	1.850	& P3L   & ELG \\
EN1\_340451	&	240.8265700	&	54.4344670	&	3	&	0	&	2.866		&	2.320	& P3    & nc,NLAGN \\
EN1\_340460	&	240.8344100	&	54.4303400	&	7	&	2	&	0.308		&	0.230	&       & SB-LINER \\
EN1\_341175	&	240.8745300	&	54.4267350	&	0	&	0	&	--		&	0.400	&       & -- \\
EN1\_340982 (s)	&	240.8745000	&	54.4215800	&	0	&	0	&	--		&	--	&       & -- \\ 
EN1\_340391	&	240.8643600	&	54.4120520	&	0	&	0	&	--		&	1.470	&       & -- \\
EN1\_341090	&	240.9015500	&	54.4095340	&	1	&	6	&	0.586		&	0.470	&       & SB \\
EN1\_339716	&	240.8470200	&	54.4039960	&	4	&	4	&	0.637		&	0.940	&       & SB \\
EN1\_339960	&	240.8671700	&	54.3996240	&	2	&	0	&	1.475		&	1.590	& P3    & BLAGN\\
EN1\_341304	&	240.9407700	&	54.3936770	&	--	&	--	&	0.000		&	--	&       & star \\
EN1\_339023	&	240.8445700	&	54.3869970	&	6	&	0	&	1.695		&	0.910	&       & BLAGN \\
EN1\_338063	&	240.8043800	&	54.3835910	&	0	&	0	&	--		&	0.990	&       & -- \\
EN1\_339817	&	240.8988300	&	54.3781200	&	0	&	0	&	--		&	0.600	&       & -- \\
EN1\_337826	&	240.8112900	&	54.3739470	&	7	&	0	&	1.756		&	1.760	& pow   & BLAGN\\
EN1\_337232	&	240.8093400	&	54.3592490	&	2	&	4	&	0.796		&	0.730	&       & SB \\
EN1\_336978	&	240.8049600	&	54.3552630	&	1	&	0	&	1.221		&	0.230	& pow   & ELG \\
\hline
\multicolumn{9}{l}{$^\ddagger$: (s) indicates serendipitous sources; }\\
\multicolumn{9}{l}{$^\ast$: pow $=$ IRAC power-law SED; P3 $=$ 5.8$\mu$m peaker; P2 $=$ 4.5$\mu$m peaker; }\\
\multicolumn{9}{l}{\ \ \ L $=$ upper limit at 8.0$\mu$m; R $=$ radio source; X $=$ X-ray source.}\\
\multicolumn{9}{l}{$^\dagger$: ELG $=$ emission lines; SB $=$ starburst diagnostics; BLAGN $=$ broad line AGN; NLAGN $=$}\\
\multicolumn{9}{l}{\ \ \ narrow-line AGN; star $=$ star; ALG $=$ absorption lines only; nc $=$ no continuum detected}\\
\multicolumn{9}{l}{\ \ \ (but emission lines yes).}\\
\end{tabular}
\normalsize
\caption{Results for second night.}
\label{tab:redshifts3}
\end{table*}

\begin{table*}[!ht]
\centering
\tiny
\begin{tabular}{l c c c c c c l l}
\hline
\hline
ID$^\ddagger$ 	& RA 	& DEC	& \multicolumn{2}{c}{Det. Lines} & $z$ & $z$ & Note$^\ast$ & Class$^\dagger$\\
SWIRE	& J2000	& J2000	& em. & abs. & spec & phot & & spec\\
\hline
EN2\_13210	&	250.416110	&	40.816822	&	1	&	0	&	0.198		&	0.450	& pow   & ELG \\
EN2\_11561	&	250.406220	&	40.777615	&	0	&	0	&	--		&	0.950	&       & -- \\
EN2\_12428	&	250.419310	&	40.801735	&	1	&	0	&	2.338		&	2.430	&       & ELG \\
EN2\_11247 (s)	&	250.416850	&	40.774560	&	4	&	0	&	0.442		&	0.410	&       & SB \\ 
EN2\_11160	&	250.420200	&	40.774158	&	0	&	0	&	--		&	1.210	&       & -- \\
EN2\_11438	&	250.426320	&	40.783157	&	2	&	3	&	0.439		&	0.510	&       & SB \\
EN2\_11745	&	250.440830	&	40.796204	&	0	&	2	&	1.561		&	1.010	&       & ALG\\
EN2\_12554	&	250.451480	&	40.817425	&	1	&	4	&	1.886		&	0.970	&       & SB \\
EN2\_10334	&	250.438280	&	40.763359	&	--	&	--	&	--		&	2.010	& P3    & -- \\
EN2\_11850	&	250.456920	&	40.805145	&	0	&	0	&	--		&	1.270	&       & -- \\
EN2\_11775 (s)	&	250.460020	&	40.805050	&	--	&	--	&	0.000		&	--	&       & star \\ 
EN2\_9745	&	250.448960	&	40.7555730	&	1	&	0	&	1.086		&	1.190	&       & ELG \\
EN2\_11091	&	250.464570	&	40.791348	&	1	&	0	&	1.946		&	2.140	& P3L   & ELG \\
EN2\_11749	&	250.480210	&	40.812843	&	1	&	7	&	0.776		&	1.150	&       & SB \\
EN2\_10010	&	250.480990	&	40.773952	&	1	&	3	&	0.673		&	0.960	&       & SB \\
EN2\_10278	&	3250.493590	&	40.785378	&	5	&	0	&	1.720		&	1.020	& pow   & BLAGN\\
EN2\_8495	&	250.4924500	&	40.7474290	&	6	&	7	&	0.362		&	0.110	&       & SB\\
EN2\_10015	&	250.508500	&	40.786121	&	0	&	7	&	0.737		&	0.710	&       & ALG\\
EN2\_9634	&	250.5140400	&	40.7809330	&	0	&	0	&	--		&	1.200	&       & -- \\
EN2\_8917	&	250.5139000	&	40.7654500	&	0	&	2	&	0.802		&	0.710	&       & ALG\\
EN2\_8175	&	250.5173800	&	40.7505040	&	2	&	0	&	0.479		&	0.330	&       & ELG \\
EN2\_9885	&	250.5391700	&	40.7966540	&	1	&	8	&	0.875		&	0.780	&       & SB \\
EN2\_9757	&	250.5478500	&	40.7976260	&	0	&	0	&	--		&	0.100	&       & -- \\
EN2\_8936	&	250.5501900	&	40.7812420	&	1	&	3	&	0.718		&	0.880	&       & SB \\
EN2\_6835	&	250.5366800	&	40.7290000	&	1	&	0	&	1.400		&	1.320	&       & ELG \\
EN2\_6702	&	250.5427900	&	40.7288020	&	1	&	3	&	0.916		&	0.980	&       & SB \\
EN2\_9202	&	250.5687000	&	40.7944600	&	1	&	7	&	0.671		&	0.590	&       & SB \\
\hline	
EN2\_172324	&	248.6213800	&	41.0597310	&	4	&	0	&	1.739		&	2.070	& P3    & NLAGN  \\
EN2\_172031	&	248.6326600	&	41.0595360	&	4	&	7	&	0.558		&	0.780	&       & SB   \\
EN2\_170417	&	248.6353800	&	41.0304950	&	0	&	0	&	--		&	1.150	&       & -- \\
EN2\_170393	&	248.6875600	&	41.0532300	&	1	&	0	&	0.912		&	1.070	&       & ELG \\
EN2\_169884	&	248.6255800	&	41.0156140	&	1	&	8	&	0.581		&	0.580	&       & SB \\
EN2\_169626	&	248.6974500	&	41.0424770	&	0	&	0	&	--		&	0.300	&       & -- \\
EN2\_168961	&	248.6883400	&	41.0261760	&	0	&	0	&	--		&	0.490	&       & -- \\
EN2\_168666	&	248.6705200	&	41.0127720	&	1	&	0	&	1.205		&	0.990	&	& ELG \\
EN2\_168355	&	248.6358200	&	40.9916650	&	1	&	0	&	0.952		&	0.490	&       & ELG \\
EN2\_167372	&	248.6506800	&	40.9797590	&	1	&	0	&	1.445		&	2.120	& P2L   & ELG   \\ 
EN2\_167270 (s)	&	248.6517500	&	40.9784900	&	0	&	0	&	--		&	1.270	&       & -- \\ 
EN2\_167232	&	248.6960300	&	40.9974060	&	5	&	8	&	0.522		&	0.290	&       & SB \\
EN2\_166853	&	248.6574900	&	40.9735030	&	1	&	4	&	0.909		&	0.890	&       & SB \\
EN2\_166134	&	248.6786000	&	40.9687960	&	1	&	0	&	1.337		&	1.470	& P2    & ELG \\
EN2\_165986	&	248.7120400	&	40.9809680	&	4	&		&	2.163		&	0.430	& P3,pow& BLAGN \\
EN2\_165843	&	248.7228500	&	40.9832570	&	0	&	0	&	--		&	0.990	&       & -- \\
EN2\_165571 (s)	&	248.7258800	&	40.9798900	&	1	&	5	&	0.797		&	0.100	&	& SB \\ 
\hline
\multicolumn{9}{l}{$^\ddagger$: (s) indicates serendipitous sources; }\\
\multicolumn{9}{l}{$^\ast$: pow $=$ IRAC power-law SED; P3 $=$ 5.8$\mu$m peaker; P2 $=$ 4.5$\mu$m peaker; }\\
\multicolumn{9}{l}{\ \ \ L $=$ upper limit at 8.0$\mu$m; R $=$ radio source; X $=$ X-ray source.}\\
\multicolumn{9}{l}{$^\dagger$: ELG $=$ emission lines; SB $=$ starburst diagnostics; BLAGN $=$ broad line AGN; NLAGN $=$}\\
\multicolumn{9}{l}{\ \ \ narrow-line AGN; star $=$ star; ALG $=$ absorption lines only; nc $=$ no continuum detected}\\
\multicolumn{9}{l}{\ \ \ (but emission lines yes).}\\
\end{tabular}
\normalsize
\begin{flushleft}
{\bf Table \ref{tab:redshifts3}.} Continued.
\end{flushleft}
\end{table*}

\begin{landscape}

\begin{table}[!ht]
\centering
\tiny
\begin{tabular}{l l l c | c c c c c}
\hline
\hline
ID    & Type  & Class  & $z$   & Mass$(\star)$		& SFR($10^8$ yr) 	& A$_V$(mean)	& A$_V$($10^8$ yr) 	& L$_{\textrm{IR}}$ 	\\
SWIRE & phot. & spec.  & spec. & $[10^{11}$M$_\odot]$	& $[$M$_\odot/\textrm{yr}]$	& $[$mag$]$	& $[$mag$]$		& $[10^{11}$L$_\odot]$ 	\\
\hline
EN1\_202261	& P2	   & ELG   &  1.339 	& 1.44 (1.41--2.36)	& 489 (171--741)	& 1.68 (1.29--1.73)	& 1.82 (1.71--1.88)	& 10.81 (8.48--12.84)	\\	
EN1\_342445	& P3L	   & ELG   &  1.917 	& 1.04 (0.91--1.84)	& 426 (268--480)	& 1.10 (0.57--1.17)	& 1.40 (0.96--1.49)	& 9.37 (6.89--12.81)	\\	
EN1\_340451	& P3	   & NLAGN &  2.866 	& 6.69 (6.12--7.29)	& 56.1 (29.6--106)	& 0.63 (0.41--0.76)	& 2.18 (1.43--2.32)	& 36.79 (25.45--50.58)	\\	
\hline
EN2\_11091	& P3L	   & ELG   &  1.946 	& 3.94 (2.37--4.20)	& 208 (197--575)	& 0.61 (0.29--0.89)	& 0.80 (0.71--1.82)	& 16.13 (10.58--20.94)	\\	
EN2\_167372	& P2L	   & ELG   &  1.445 	& 2.62 (2.08--2.77)	& 86.9 (25.4--192)	& 0.59 (0.07--0.93)	& 1.34 (0.35--1.80)	& 4.98 (0.96--12.77)	\\	
EN2\_166134	& P2	   & ELG   &  1.337 	& 3.40 (3.05--3.86)	& 17.9 (9.90--200)	& 0.79 (0.50--0.89)	& 1.65 (1.20--2.02)	& 9.06 (5.82--12.41)	\\	
\hline
LH\_572243	& P2,X     & NLAGN &  1.820 	& 0.91 (0.85--0.92)	& 34.3 (27.7--91)	& 0.82 (0.79--0.90)	& 2.09 (1.99--2.28)	& 9.27 (9.04--10.51)	\\      
\hline
\end{tabular}
\normalsize
\caption{Results of spectro-photometric fitting for sources with no AGN component.
The five columns on the right contain the stellar masses of galaxies, their ongoing (within the last $10^8$ yr)
star formation rate, two values of the intrinsic extinction,
and the IR ($8-1000$ $\mu$m restframe) luminosity produced by the 
ongoing starburst, estimated assuming an M82-like template. The two $A_V$ values are computed 
by averaging over the whole galaxy life 
and over the last $10^8$ yr (representing the amount of dust affecting the ongoing burst). In parenthesis
the 3$\sigma$ ranges derived from the exploration of the parameter space are reported.}
\label{tab:fit_sb}
\end{table}

\begin{table}[!ht]
\centering
\tiny
\begin{tabular}{l l l c | c c c c c c | c c c | c | c c}
\hline
\hline
ID 	& Type	& Class  & $z$   & \multicolumn{6}{|c|}{Torus} & \multicolumn{3}{|c| }{Stellar population} & Total & \multicolumn{2}{c}{\% AGN} \\
SWIRE 	& phot.	& spec.  & spec. & $\frac{R_{out}}{R_{in}}$ & $\Theta^\dagger$ & $\tau_{9.7}$ & $\Psi^\ddagger$ & L$_{BH}$ & $\alpha$ & age & E(B--V) & Mass & $L_{IR}$ & 8.0$\mu$m & IR \\
& & & & & $[$deg$]$ & & $[$deg$]$ & $[10^{45}$  L$_\odot]$ & index$^\ast$ & $[10^6$ yr$]$ & $[$mag$]$ & $[10^{10}$M$_\odot]$ & $[10^{11}$ L$_\odot]$ & & \\
\hline
EN1\_202260	& P2,X     & BLAGN &  1.545 & 100--300 & 100--140 & 0.6--2.0  & 0--30 & 0.79--1.58 & -1.0 & 50--300 & 0.2--0.4 & 3.98--10.0 & 4.23--7.05 & 53--71\% & 10--62\% \\   
EN1\_282078	& P3	   & BLAGN &  1.685 & 20--100  & 60--100  & 6.0--10.0 & 0--60 & 1.58--3.16 & -1.0 & 50--100  & 0.6--0.8 & 3.98--6.31 & 10.4--23.7 & 53--65\% &  2--40\% \\   
EN1\_279954	& P3,pow   & BLAGN &  2.409 & 30--100  & 60--140 & 0.6--3.0  & 0--50 & 1.58--3.16 & -0.5 & 10--100  & 0.6--1.0 & 2.51--15.8 & 15.4--27.1 & 49--63\% &  5--31\% \\   
EN1\_339960	& P3	   & BLAGN &  1.475 & 30--100  & 60--140 & 0.6--1.0  & 0--50  & 0.20--0.32 & -1.0 & 50--500 & 0.2--0.6 & 3.98--25.1 & 3.83--5.50 & 39--49\% & 10--48\% \\   
\hline
EN2\_275226	& P2	   & BLAGN &  1.710 & 30--100  & 100--140 & 1.0--6.0  & 0--20  & 0.32--1.58 & -0.5 & 10--50  & 0.6--0.8 & 1.58--2.51 & 15.3--26.1 & 37--54\% &  1--20\% \\   
EN2\_273717	& P3	   & BLAGN &  1.800 & 30--100  & 100--140 & 0.3--3.0  & 0--20  & 1.00--2.51 & -0.5 & 10--50  & 0.6--0.8 & 1.00--2.51 & 26.4--38.2 & 48--62\% &  2--24\% \\   
EN2\_172324	& P3	   & NLAGN &  1.739 & 30--100  & 60--140  & 0.3--2.0  & 40--90 & 1.00--3.16 & -1.0 & 50--100 & 0.4--0.6 & 15.8--25.1 & 16.4--23.4 & 32--45\% &  3--39\% \\   
EN2\_165986	& P3,pow   & BLAGN &  2.163 & 20--100  & 80--140  & 3.0--10.0 & 30--50 & 3.16--10.0 & -1.0 & 10--100  & 0.2--0.6 & 1.00--10.0 & 11.5--18.6 & 64--80\% & 20--53\% \\   
\hline
LH\_574364	& P3,R     & NLAGN &  1.474 & 100--300 & 100--140 & 0.3--3.0  & 30--70 & 1.58--1.99 & -0.5 & 50--100 & 0.4--0.8 & 2.51--6.31 & 7.42--9.01 & 55--63\% & 30--77\% \\   
\hline
\multicolumn{16}{l}{$^\dagger$: the aperture angle $\Theta$ is computed starting from the equatorial plane,
and it is doubled, accounting for equatorial symmetry.}\\
\multicolumn{16}{l}{$^\ddagger$: the viewing angle $\Psi$ is computed starting from the polar axis.}\\
\multicolumn{16}{l}{$^\ast$: source power law slope in UV-optical-IR range ($\lambda\, L[\lambda]\propto\lambda^\alpha$).}\\
\end{tabular}
\normalsize
\caption{Results of SED fitting for IR-peakers requiring an AGN component. The fit consists in the 
combination of a simple stellar population and a torus model \citep{fritz2006}. The 3$\sigma$ ranges for 
the main geometrical and dust properties of the torus are reported, as well as stellar population ages, extinctions and masses.}
\label{tab:fit_agn}
\end{table}

\end{landscape}

\begin{table*}[!ht]
\centering
\begin{tabular}{l r r}
\hline
\hline
  & No. & \% \\
\hline
Observed slits		& 235	& -- \\
Original targets	& 233	& -- \\
Serendip. targets	& 68	& -- \\
Total targets		& 301	& -- \\
Tot. measured redshifts	& 174/301 & 58\% \\
Serendip. redshifts	& 35/68	& 52\% \\
SWIRE redshifts		& 150	& -- \\
\hline
Absorption line glxs	& 7/150 & 5\% \\
Emission line glxs	& 122/150 & 81\% \\
Starbursts		& 39/150 & 26\% \\
Type-2 AGNs		& 5/150 & 3\% \\
Type-1 AGNs		& 17/150 & 11\% \\
$z>2.5$ QSOs		& 3/150 & 2\% \\
\hline
Obs. IR-peakers		& 35	& -- \\
Obs. 4.5$\mu$m-peakers	& 8/35	& 23\% \\
Obs. 5.8$\mu$m-peakers	& 27/35	& 77\% \\
IR-p. with redshift	& 16/35	& 46\% \\
4.5$\mu$m-p. with redshift & 6/8 & 75\% \\
5.8$\mu$m-p. with redshift & 10/27 & 37\% \\
AGN IR-peakers		& 11/16	& 69\% \\
Type-1 AGN IR-peakers	& 7/11	& 64\% \\
Type-2 AGN IR-peakers	& 4/11	& 36\% \\
Type-1 AGN 4.5$\mu$m-peakers	& 2/7	& 29\% \\
Type-1 AGN 5.8$\mu$m-peakers	& 5/7	& 71\% \\
Type-2 AGN 4.5$\mu$m-peakers	& 1/4	& 25\% \\
Type-2 AGN 5.8$\mu$m-peakers	& 3/4	& 75\% \\
\hline
\end{tabular}
\caption{Summary of results.}
\label{tab:summary}
\end{table*}


\begin{landscape}

\begin{figure}[!ht]
\centering
\rotatebox{-90}{\includegraphics[width=0.90\textwidth]{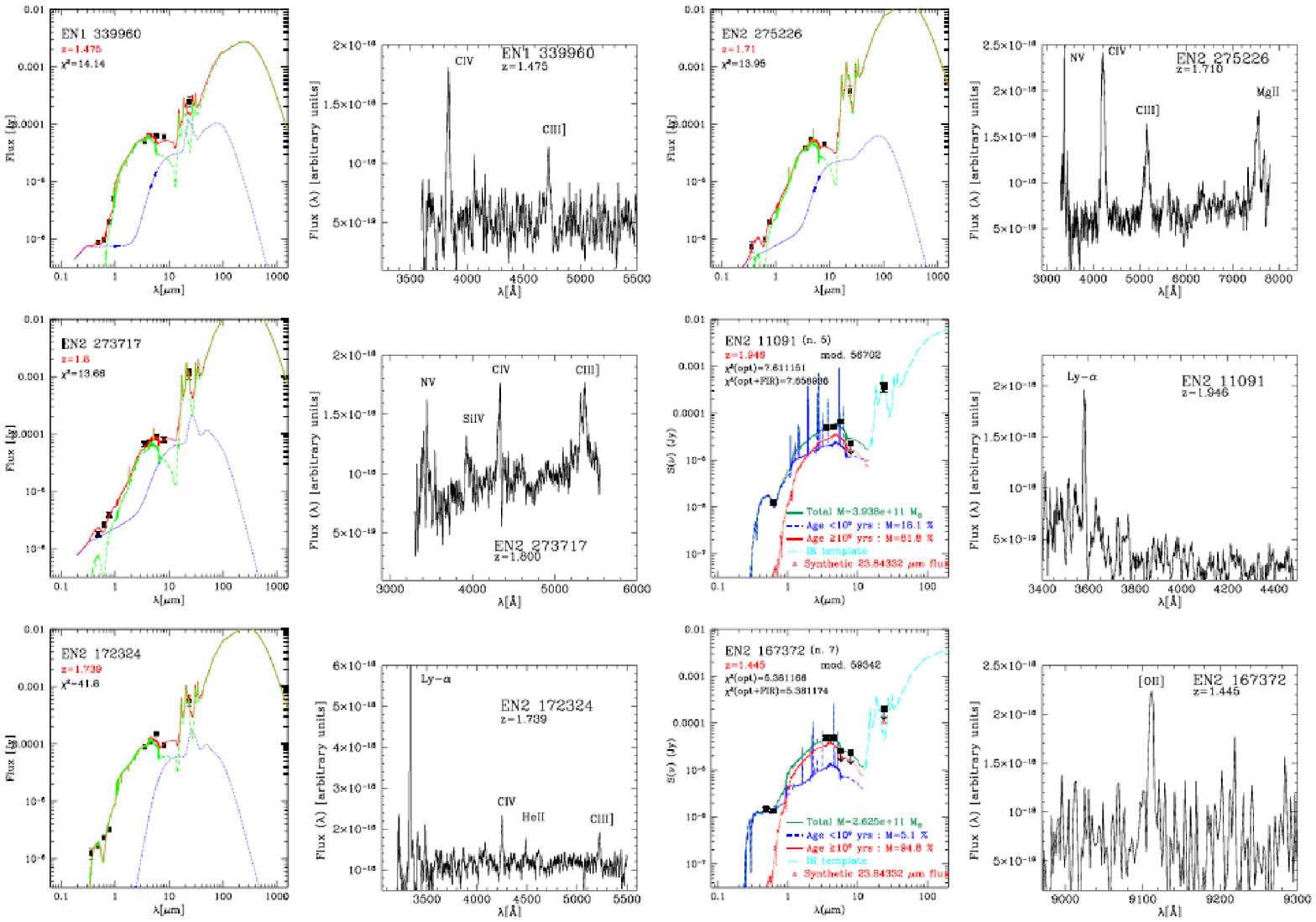}}
\begin{flushleft}
{\bf Fig. \ref{fig:ib_seds1}.} Continued.
\end{flushleft}
\end{figure}

\end{landscape}

\begin{landscape}

\begin{figure}[!ht]
\centering
\includegraphics[height=0.60\textwidth]{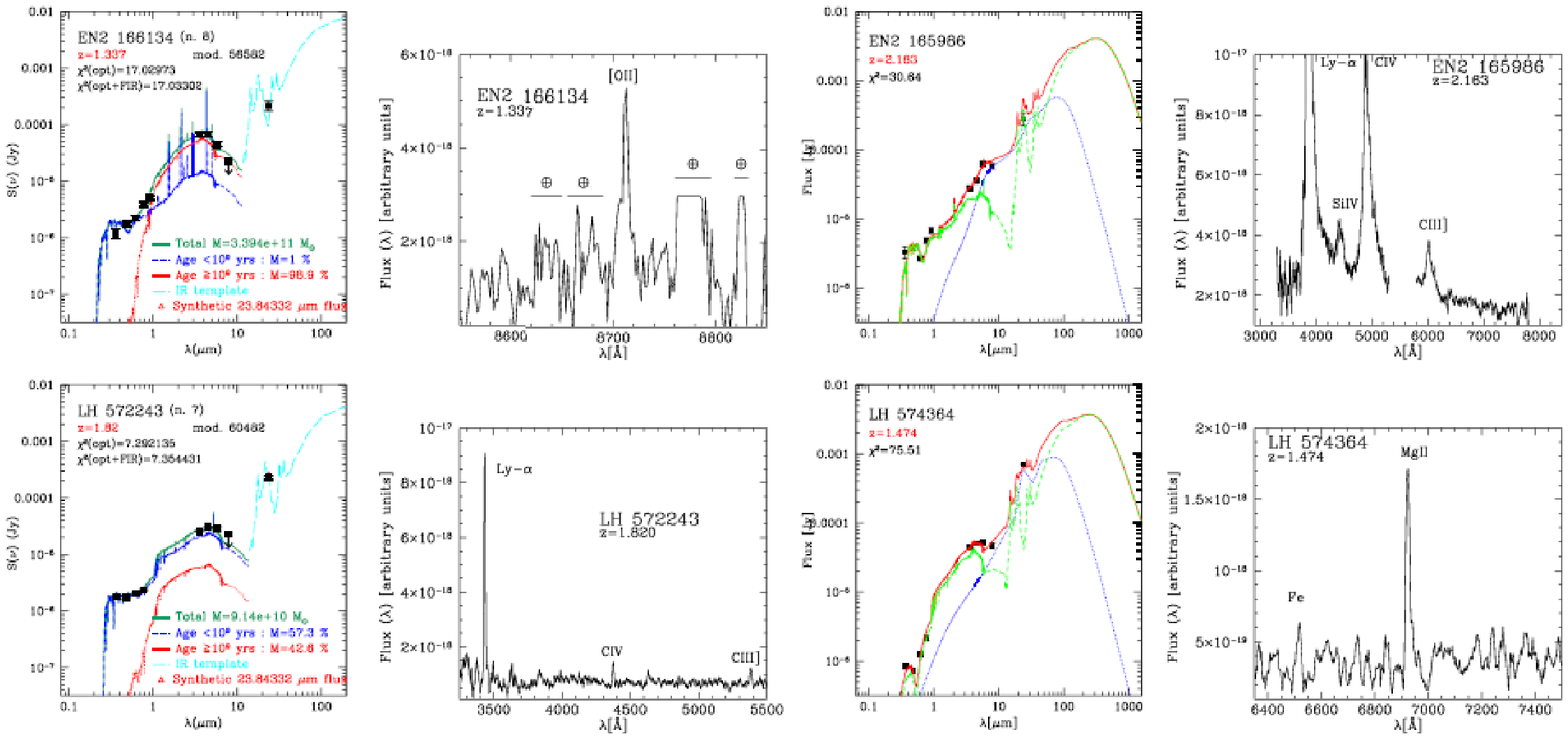}
\begin{flushleft}
{\bf Fig. \ref{fig:ib_seds1}.} Continued.
\end{flushleft}
\end{figure}

\end{landscape}

\end{document}